%% file: main_spectral_fitting.tex
\documentclass[]{agujournal2019}
\usepackage{url} %
\usepackage[inline]{trackchanges} %
\usepackage{soul}
\usepackage{longtable}
\usepackage{booktabs}

\newcommand{\fc}{\ensuremath{f_\mathrm{c}}}

\newcommand{\Azerosq}{\ensuremath{A^2_{\scriptscriptstyle 0}}}

\newcommand{\MW}{\ensuremath{M_{\mathrm{w}}}}
\newcommand{\QS}{\ensuremath{Q_{\mathrm{S}}}}
\newcommand{\SPratio}{\ensuremath{\Delta A_{\mathrm{PS}}}}
\newcommand{\flow}{\ensuremath{f_{\mathrm{\;low}}}}
\newcommand{\fhigh}{\ensuremath{f_{\mathrm{\;high}}}}
\newcommand{\tstarP}{t^*_{\mathrm{P}}}
\newcommand{\tstarS}{t^*_{\mathrm{S}}}
\newcommand{\vS}{\ensuremath{V_{\mathrm{s}}}}
\draftfalse

\journalname{JGR: Plonks}
\begin{document}

\title{Discrimination of tectonic, swarm and impact-related marsquakes using spectral characteristics}

\authors{Simon C. St\"ahler\affil{1,2}, Savas Ceylan\affil{1}, Nikolaj Dahmen\affil{1}, John Clinton\affil{3}, Domenico Giardini\affil{1}}
\affiliation{1}{Institute of Geophysics, ETH Zurich}
\affiliation{2}{ETH Space, ETH Zurich}
\affiliation{3}{Swiss Seismological Service, ETH Zurich}

\correspondingauthor{Simon C. St\"ahler}{mail@simonstaehler.com}

\begin{keypoints}
\item We build a simple spectral model to explain all observed marsquakes
\item We find that marsquakes form 3 distinct families - tectonic, swarm and impacts
\item We derive new impact size predictions and estimate Mars' seismicity
\end{keypoints}

\begin{abstract}
The NASA InSight mission observed over 2000 marsquakes in the course of its three year mission. These quakes varied in magnitude between 1.5 and 4.5, as well as in spectral content. We present a simple framework to describe the spectral characteristics of all observed marsquakes, based on source process; propagation through the mantle or crust; and local, receiver-side amplification. We assign to each quake an objective measure of its amplitude, as well as the spectral decay created by the duration of the rupture and the dampening of high frequencies due to visco-elastic attenuation. Together, this allows us to obtain characteristic patterns of the whole marsquake dataset, e.g. in terms of event magnitudes, source size, and - for quakes caused by meteoritic impacts - crater size. We show that a significant fraction of all marsquakes - the high-frequency quakes - form a swarm that is likely not caused by tectonic processes in rocks. 
Our analysis allows separation of the whole marsquake catalogue into three event classes, of tectonic quakes, meteoritic impacts, and swarm events. We finally conclude that the largest marsquake, S1222a, most likely belongs to the group of meteoritic impacts.
\end{abstract}

\section*{Plain Language Summary}
Over the duration of the InSight mission, very different kinds of marsquakes were observed and initially sorted into rather technical categories. We present a simplified framework of families directly related to the source process: Tectonic quakes due to crustal strain, Swarm quakes of a so-far not understood mechanism, and meteoritic impacts. We show that these three clusters form separate groups in a plot of event size over duration (or magnitude over corner frequency). From this observation, we follow that the largest marsquake observed so far, S1222a, most likely is caused by a meteoritic impact. We also show a complex source-side resonance effect in most impact-generated marsquakes. We finally use our dataset to quantify the overall number of marsquakes and find it to be between the Moon and the pre-mission predictions that stood at the beginning of the InSight mission concept.

\section{Introduction}
An early observation of the InSight seismic observation campaign on Mars was that Marsquakes come in multiple distinct flavors \cite{giardini_seismicity_2020, banerdt_initial_2020}. The first significant marsquake, observed on Sol 128 of the mission, was characterized by a long coda, i.e. ringing seismic energy for several minutes and a relatively high corner frequency $\fc$ of more than 5~Hz. No clear seismic phases were observable and only slope breaks in the envelope of the signal hinted at a secondary, S-wave arrival \cite{lognonne_constraints_2020}, from which a distance was estimated. Later, on Sol 173, a seismic event was observed that was more similar to a teleseismic earthquake, with several clearly distinct seismic phases, and energy being limited to below one Hz. This led the InSight core service team tasked with detecting and classifying marsquakes, the \textit{Marsquake Service} (MQS), to introduce two classes of Marsquakes: \textit{Low-frequency marsquakes} (LF) and \textit{High-frequency marsquakes} (HF) \cite <for details, see>{clinton_marsquake_2021}. In the months that followed, on Sol 218 (2019-07-09), another type of marsquake was observed, which showed two clear energy arrivals of narrow-band high-frequency energy, separated by about 5 minutes. Initially, this was identified as a doublet marsquake, but over the next weeks other very similar events occurred. This led to the identification of the two bumps as guided reverberations of P- and S-waves in the upper crust \cite{van_driel_high-frequency_2021}. This new event type had significant energy above 1 Hz, but not quite as high as the first marsquake on Sol 128. This led to the introduction of a new class, the \textit{Very High frequency} (VF) marsquakes, in which S0128a was placed, while S0218a-like events now populated the HF class. The InSight seismic data show a strong spectral peak around 2.4 Hz \cite{dahmen_resonances_2021, hobiger_shallow_2021}, which is excited by all marsquakes of the HF and VF class \cite{van_driel_high-frequency_2021}, to the extent that weak HF events are sometimes only visible as an increase of energy around 2.4 Hz. Thus, these weak HF quakes were called \textit{ 2.4 Hz} events. On Sol 235, the last member of the marsquake bestiary entered: An event that had clear P- and S-waves, but also excited the 2.4 Hz resonance. To distinguish these types of events, the family of \textit{Broadband} (BB) events was introduced. A separate observation was that around dusk and dawn, very short events with a duration of less than 30 seconds and a frequency content above 5 Hz occurred regularly \cite{dahmen_super_2020}. Although a conclusive explanation for these has not yet been found, an association with thermal cracking of nearby rocks, similar to thermal quakes in the Apollo lunar seismic data set, is plausible \cite{duennebier_thermal_1974, civilini_detecting_2021}. This class of quakes was termed super-high-frequency quakes (SF) and will not be further discussed here.

This initial classification schema held up surprisingly well over the duration of the InSight mission, although it was observed that some HF events did include energy straying below 1 Hz. More significant cracks in the schema began to show in September and December 2021, when two large meteoritic impacts hit Mars \cite{posiolova_largest_2022} and created broadband seismic signals with high corner frequencies, extended coda, like a VF or HF event, but also a time difference between P and S arrival times indicative of wave propagation through the deep mantle. The identification of the main arrivals as mantle-going waves is consistent with the travel times expected given the observed location of the impact craters \cite{duran_observation_2022}, yet the long coda suggests strong scattering near the surface. These two large events were termed Wideband events (WB), since they contained observable energy over the full range of the InSight seismometer, from 0.1 Hz to up to 10 Hz (and above, in some cases). Surprisingly, a range of other wideband events happened late in the mission, so on Sols 1102 (S1102a) and 1133 (event S1133c). A grand finale of the mission was the S1222a event, the largest ever observed marsquake, with a magnitude of 4.7 \cite{kawamura_s1222alargest_2023}, which allowed to study surface waves over a broad frequency range, hence improved our understanding of the Martian crust, and warranted its own special issue in JGR planets. The large size and unusual source region triggered a search for an impact crater, which was unsuccessful so far \cite{fernando_tectonic_2023}.

The various event classes were only gradually associated with different source processes. Initial analysis of secondary phases and focal mechanisms \cite{brinkman_first_2021, jacob_seismic_2022} indicated source depths of 20-50 km for the LF and BB events and polarization analysis of body waves indicated a location of most of the events near the Cerberus Fossae graben system in Southern Elysium Planitia \cite{giardini_seismicity_2020, zenhausern_lowfrequency_2022}. Given a lack of clear polarization, the HF and VF events could initially not be clearly located, but the long coda was interpreted to be the result of a shallow source within a crustal layer that traps waves to some extent, similar to waves trapped in strongly layered oceanic crust on Earth \cite{kennett_high-frequency_2013}. \citeA{stahler_tectonics_2022} proposed a tectonic model in which a large scale extensional stress field around Cerberus Fossae can explain both HF and LF/BB events: LF/BB events occur deeper in the crust, where intruding magma weakens the rock, while HF/2.4 Hz events occur in the surficially visible grabens. This model was the first to take into account the spectral characteristics, specifically the difference in corner frequency \fc. Specifically, it showed that the lack of high frequency energy in the signal cannot be explained by attenuation. Given the small magnitudes ($\MW<4.6$) and distances above 25\textdegree{}, we can treat all marsquakes as point sources.

At this point, a small excursion to Earth is warranted \cite{warren_investigating_2000}. For shallow teleseismic earthquakes, P-waves typically show energy up to a few Hz, with flat spectra in displacement below about 1 Hz (termed the corner period). S-waves are significantly depleted in high-frequency energy and their spectra begin to drop at lower frequencies, at about 0.2 Hz. This is a combination of two effects: 1. Larger earthquakes have a longer rupture process and thus radiate less coherent energy at high frequencies. This spectral shape can be described by a power-law 
\begin{equation}
    S(f) =  A_0  / (1 + (f/\fc)^n),
\end{equation}
where $S(f)$ is the spectral amplitude at frequency $f$, $A_0$ is the peak amplitude at long period, and $\fc$ is the corner period. The exponent $n$ varies between $2<n<3$, depending on the earthquake source model \cite{brune_tectonic_1970, madariaga_dynamics_1976}. 2. Viscoelastic attenuation, $Q_{\mu}$, damps high-frequency energy during propagation. This effect is significantly stronger for S- than for P-waves. Only for earthquakes larger than magnitude 6 or 7 is the source effect typically visible in S-wave spectra \cite{shearer_introduction_2019}.

\begin{figure}
    \centering
    \includegraphics[width=0.95\textwidth]{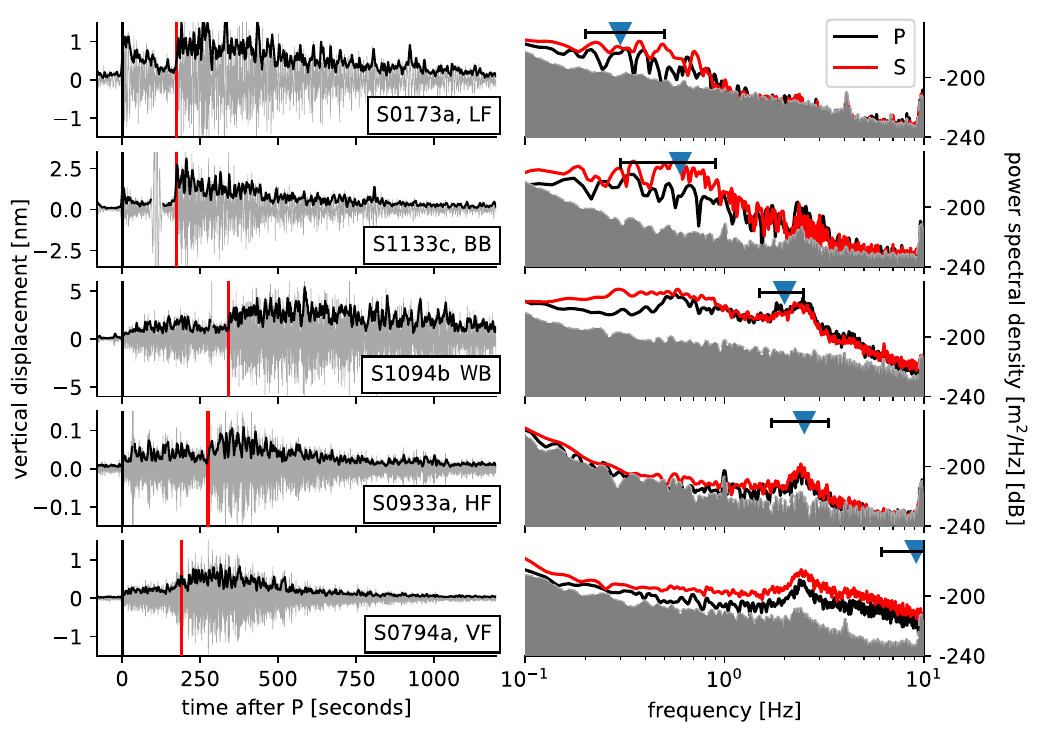}
    \caption{Spectral and time-series character of the five event types encountered during the mission. The left columns shows time series of vertical displacement, with P and S-arrivals marked in black and red respectively. The black solid line shows a 5-second smoothed envelope of the time series. The right column shows displacement spectra on the vertical component, with the same colors. The grey shaded area marks the pre-event noise. Note that despite these five events being among those with highest SNR in their respective class, the signal exceeds this noise level only in a relatively narrow frequency band. The blue triangles mark our estimate of the corner frequency $\fc$, with associated error bar (see sect. \ref{sec:model}) . Since the attenuation affects the spectra as well, for event S0794a, the spectral energy starts decaying already at frequencies below \fc.
    }
    \label{fig:overview}
\end{figure}

Figure \ref{fig:overview} shows examples of timeseries and displacement spectra of all marsquake event types mentioned above. As can be seen, all spectra can be explained by a classical model of a flat part below a corner frequency and a decay proportional to $f^2$ plus attenuation at higher frequencies. The different interpretations throughout the mission can be attributed 
a) to the relatively low signal-to-noise ratio of most events, which typically only surpass the ambient noise in a narrow frequency range, sometimes less than two octaves (e.g. the HF event S0933a is only visible between 1 and 4 Hz, despite having the highest SNR of its class).
b) to the local amplification around 2.4 Hz, which leads to somewhat unusual spectra for low-SNR events, that may only exceed the noise around this specific frequency.

On Mars, the S-wave spectrum of LF events could have been easily explained by attenuation after propagation over the 1500 km epicentral distance to Cerberus Fossae. However, the P-wave spectra of LF marsquakes show almost the same spectrum as the S-waves, with similarly depleted high frequency content, which is incompatible with attenuation. In the absence of strong scattering (for the LF/BB events at least), this hints at a slow source process, which was interpreted by \citeA{stahler_tectonics_2022} as indicative of warm, weakened material.

The location of HF events warrants further discussion. \citeA{stahler_tectonics_2022} has interpreted them as shallow events of Cerberus Fossae, based on a wave propagation model, where energy is mostly trapped in the crust and therefore travels very slowly \cite{giardini_seismicity_2020, van_driel_high-frequency_2021}. However, we want to correct this interpretation for two reasons. First, more detailed numerical modeling could not replicate the crust-only propagation model proposed in \citeA{van_driel_high-frequency_2021} without assuming a very strong attenuation in the upper mantle. This was further supported by the association of a fresh meteoric impact crater in Cerberus Fossae with the VF event S0794a \cite{charalambous_new_2025, bickel_new_2025}. Second, as shown in a recent re-analysis \cite{dahmen_analysis_2026}, the back-azimuth determined for some HF events in \citeA{stahler_tectonics_2022} is likely a spurious signal in a small subset of events that cannot be reproduced. \cite{dahmen_analysis_2026} recomputed the distance of all HF events with the same mantle-based wave propagation model that is used for all other event classes, resulting in larger epicentral distances between 40 and 50\textdegree{}, identified and discussed potential source regions and labeled the whole set as a swarm, although so far there is no clear candidate. 

The objectives of this study and the scope of the paper are as follows: to conduct a systematic manual reexamination of the spectral content present in all significant marsquakes. This involves revisiting and potentially revising the classification of existing event types and progressing towards a novel interpretation of marsquake phenomena. This paper serves as a companion to the work by \citeA{dahmen_analysis_nodate}, which employs machine learning to enhance the identification of high-frequency (HF) events, improve their localization, and redefine them using mantle models for distance estimation. As a result of this analysis, the majority of HF events that are likely to have a common source region and a similar seasonal event rate are labeled as swarm events. In contrast, HF events with a different source region and seasonality are associated with low signal-to-noise ratio (SNR) versions of VF events. Furthermore, this study is situated within the broader context of a comprehensive review of the entire marsquake catalogue, which will be elaborated in a forthcoming paper. 

\section{Method}
\subsection{Spectral estimation}
\begin{figure}
    \centering
    \includegraphics[width=0.95\linewidth]{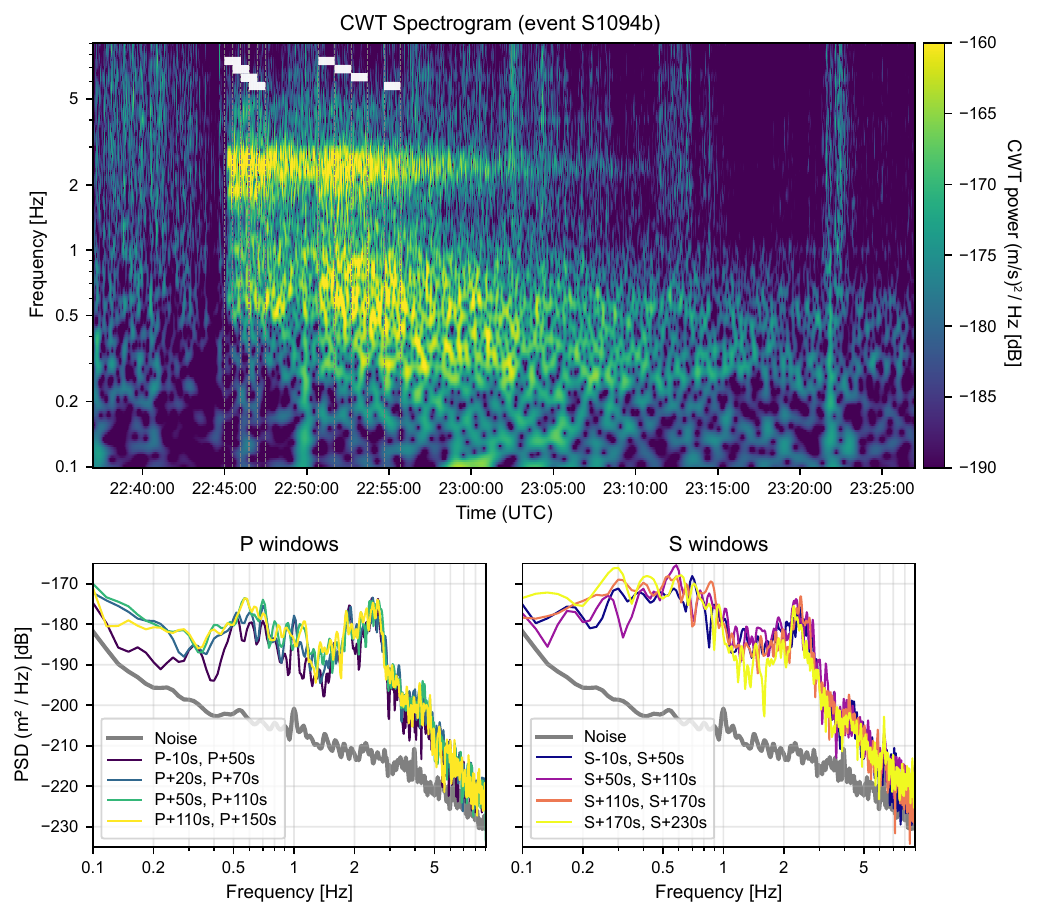}
    \caption{Vertical component spectrogram of event S1094b, caused by a meteoritic impact \cite{posiolova_largest_2022}. The bottom two panels show P- and S-wave spectra in several 60 second time windows following the direct arrival. The spectral shape stays consistent throughout the inital 150 and 230 seconds after arrival of the phase, indicating an equipartitioning of seismic energy with no significant viscoelastic attenuation.}
    \label{fig:spectrogram_S1094b}
\end{figure}

Spectra were computed from displacement time series recorded by the very broadband main sensor (VBB) \cite{lognonne_seis_2019}. For the vast majority of events, these were recorded with 20 samples per seconds (sps), i.e. up to a Nyquist frequency of 10 Hz. Where available, recordings with 100 sps, either on the SP-short period sensor or - late in the mission - on the VBB sensor were used to investigate spectra above 10 Hz. 
Note that seismic noise on Mars is highly time-dependent, with wind gusts and temperature-driven instrument glitches \cite{scholz_detection_2020} creating disturbances which regularly exceed typical marsquake signal energy. Therefore, great care was taken to identify signal windows that are free of such disturbances and noise time windows that are characteristic of the noise floor during the event. As described in \citeA{clinton_marsquake_2021, ceylan_companion_2021}, the goal was to identify time windows of at least 40 seconds, if possible. The spectra were computed using Welch's method with a window length of 20 seconds, padded by a factor of 2, with 50\% overlap. 
Figure \ref{fig:spectrogram_S1094b} shows spectra of several 60 second time windows through the initial coda of one strong seismic event, the meteoritic impact S1094b. The spectra show no variation, indicating a stable spectral shape. This means that our spectra computed in the early coda are representative of the whole event.

As a prerequisite to the spectral analysis, the frequency band above the background noise was determined for each of the P and S-wave spectra. The highest and lowest frequency were tracked.  
\clearpage
\subsection{Spectral fitting model} 
In this study, we introduce a straightforward model designed to describe the spectral features of all marsquakes, employing seven independent parameters, $\Azerosq, \SPratio , \fc, t^*, f_w, A_{2.4}, f_{2.4}$. The foundational premises are as follows:
\begin{enumerate}
  \item The spectral energy recorded during the event $A^2(f)$ is the sum of the energy of the event itself $A_{\textrm{\tiny{event}}}^2(f)$ and noise $A_{\textrm{\tiny{noise}}}^2(f)$. The noise spectrum can be well estimated from a time window before the quake. 
  \item All marsquakes can be described by a Brune-like source of a flat displacement spectrum \Azerosq{} up to a corner frequency \fc, and a decay proportional to $f^{-n}$ above (we use n=2).
  \item We model both P- and S-spectra of the source with the same corner frequency, but a constant offset \SPratio.
  \item In the relatively narrow frequency range in which we observe the signal above ambient noise, the attenuation can be described by a frequency-independent quality factor $Q$. 
  \item The intrinsic attenuation quality factor $Q$ is dominated by $Q_{\mu}$, so that: \begin{equation}
      Q_{\textrm{\small{P}}}=9/4Q_{\textrm{\small{S}}}
  \end{equation} for a Poisson medium and 
  \begin{eqnarray}
      \tstarP &=& T_{\textrm{\small{P}}}/Q_{\textrm{\small{P}}} \nonumber \\
      &=& (T_{\textrm{\small{S}}}/\sqrt{3}) / (9/4Q_{\textrm{\small{S}}})\\
      &\approx& \tstarS/3.9 \nonumber
  \end{eqnarray}
  For brevity, we refer to $\tstarS$ as $t^*$.
  \item The observed spectral ratio of horizontal vs. vertical energy (H/V ratio) on both phases is shaped strongly by the local subsurface. 
  \begin{enumerate}
  \item One aspect is the vertical amplification around 2.4 Hz \cite{hobiger_shallow_2021}. We test whether it is constant for all events, by fitting the spectral peak with parameters $f_w, A_{2.4}, f_{2.4}$ (see below).
  \item The the amplification of energy on the horizontal component above 4 Hz \cite{carrasco_empirical_2023} is treated equally for all events (see next section). 
\end{enumerate}
\end{enumerate}

\subsection{H/V ratios}
\label{sec:HV}
Before fitting, we use our complete dataset to review the H/V ratio curve, which \citeA{carrasco_empirical_2023} had based on a subset of data available at the time. Compared to this earlier analysis, we now have a set manually reviewed frequency band in which the signal exceeds the noise. We find a good match between all event types up to 10 Hz (fig. \ref{fig:HV_curve}). Table \ref{tab:HV} in appendix shows the observed H/V ratios between 0.3 and 9.2 Hz. Due to the limited instrument response close to the Nyquist frequency of 10 Hz, this is the highest value we confidently identify. Note that compared to \citeA{carrasco_empirical_2023}, we find an additional peak in H/V at 3.9 Hz. This peak is present in some of their spectra (see fig. 5 in their paper), but was smoothed out in their published median curve. As reported by various authors, \cite{dahmen_resonances_2021, carrasco_empirical_2023}, the H/V curves of the two sensors are not consistent above 10 Hz, so we restrict our analysis to range below.
\begin{figure}
    \centering
    \includegraphics[width=\textwidth]{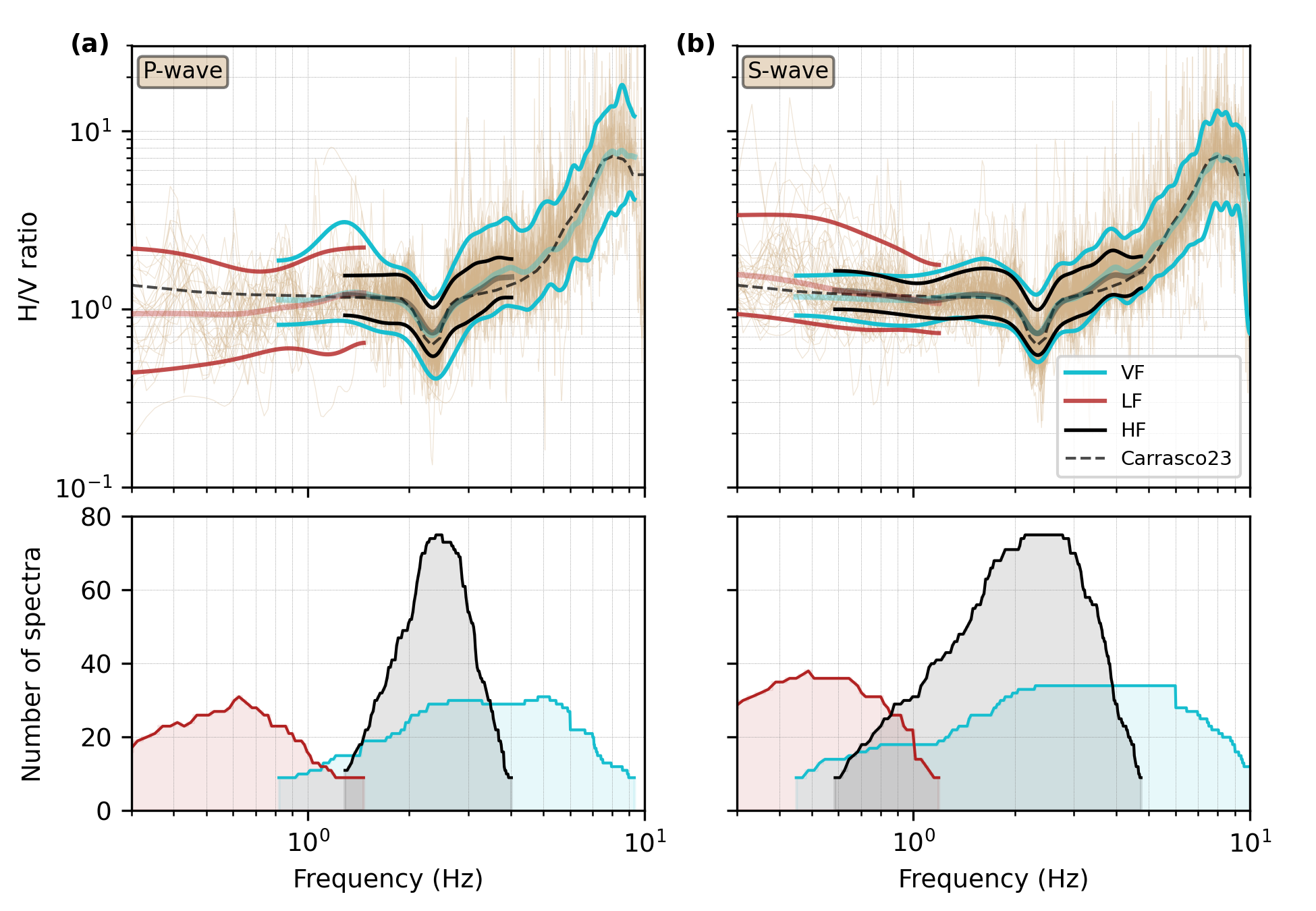}
    \caption{Ratio of horizontal and vertical energy (H/V ratio) for different event types. Yellow lines are individual events, strong lines are median and 5th and 95th percentiles of different event types: LF and BB events, VF events, and HF events. While the event types cover significantly different frequency ranges, the H/V curve in the overlapping parts is consistent, showing that the excessive horizontal energy of the VFs above 5 Hz is a common effect for all events which exceed the background noise at these frequencies. The dashed line is the smoothened H/V curve described by \citeA{carrasco_empirical_2023}.}
    \label{fig:HV_curve}
\end{figure}
The H/V analysis extended to all events of sufficient SNR demonstrates that there is no specific amplification of energy on the horizontal component for the VF events only, as had been previously assumed \cite{clinton_marsquake_2021}. Every marsquake signal that has energy above noise for frequencies $>5$ Hz shows an H/V ratio $>1$, i.e. an excess of energy on the horizontal component compared to the vertical. This explanation is parsimonious, and requires no extra mechanism on source or propagation side, which would have been only triggered by some event types, specifically VF events. 

\subsection{Model}
\label{sec:model}
\begin{figure}
    \centering
    \includegraphics[width=0.95\textwidth]{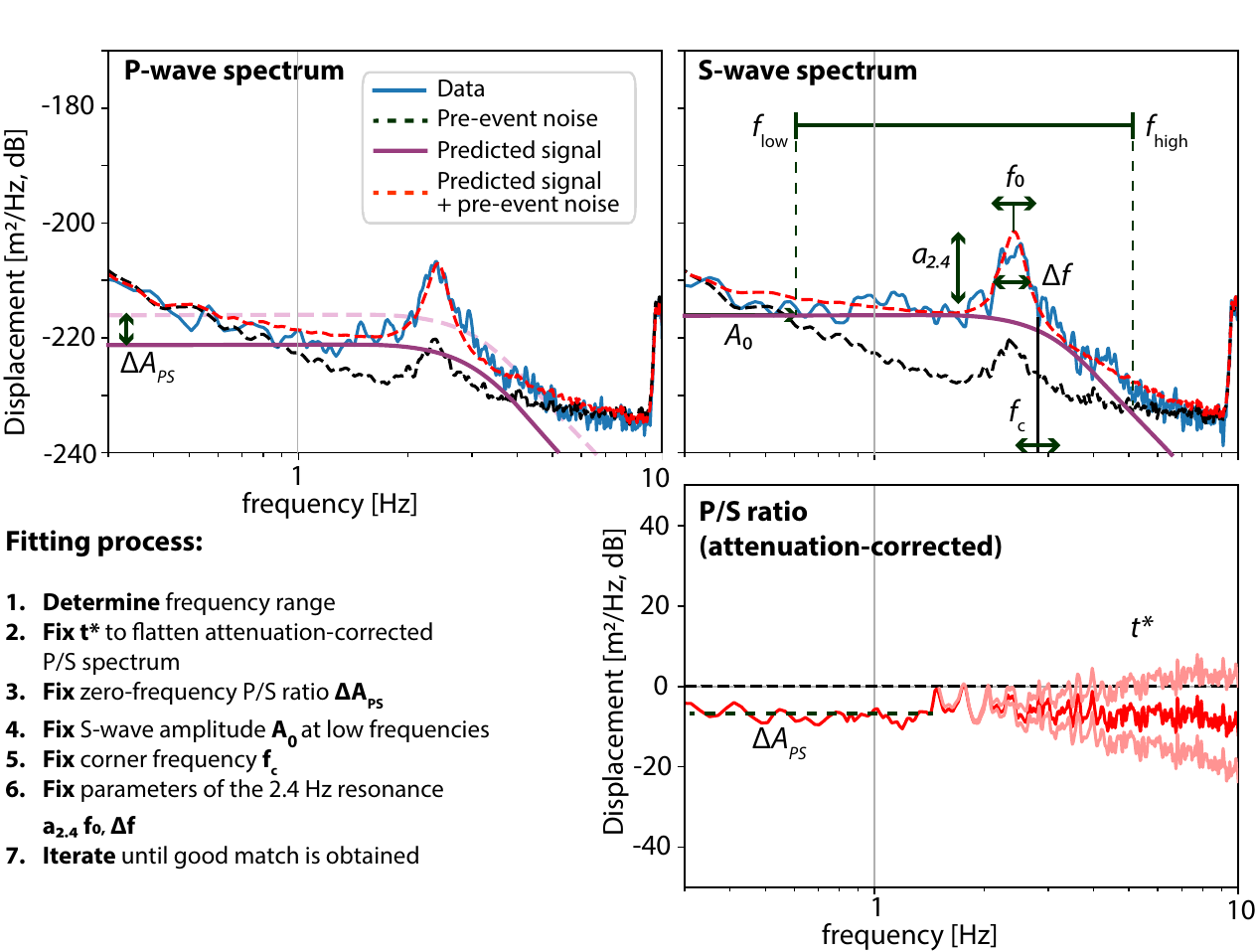}
    \caption{Free parameters in the spectral fitting process}
    \label{fig:spectral_fitting}
\end{figure}

The P and S-spectra $A^2_{\textrm{P}}(f), A^2_{\textrm{S}}(f)$, therefore, can be parametrized as
\begin{eqnarray}
    A^2_{\textrm{S}}(f) & = & \left(A_0 \times s(f) \times a_{\textrm{att,S}}(f) \times a_{\textrm{2.4}}(f)\right)^2 + A^2_{\textrm{noise}}(f) \textrm{ and} \\
    A^2_{\textrm{P}}(f) & = & \left(A_0 \times \SPratio \times s(f) \times a_{\textrm{att,P}}(f) \times a_{\textrm{2.4}}(f)\right)^2 + A^2_{\textrm{noise}}(f)\textrm{, where} \\
    s(f) & = & \frac{1}{1 + (f/f_c)^n} \\
    a_{\textrm{att,S}}(f) & = & \exp{\left(-\pi f t^* \right)} \\
    a_{\textrm{att,P}}(f) & = & \exp{\left(-\pi f t^*/3.9 \right)} \\
    a_{\textrm{2.4}}(f)  & = & 1 + \frac{a_{2.4}}{1+\left(\frac{f-f_0}{\Delta f/2}\right)^2} 
\end{eqnarray}
The free parameters for the source are the zero-frequency amplitude, $A_0$ of the S-wave; the P/S energy ratio at long periods, $\SPratio$; and the corner frequency of the source \fc. The free parameter for the attenuation $a_{\textrm{att,i}}(f)$ is described by the $t^*$-value of the S-wave. This paper focuses on a quantitative way to describe Marsquake spectra, so we will not try to invert these $t^*$ for attenuation. This would require a propagation model, which does not yet exist for the HF family events. The local amplification is described by a Lorenz curve $a_{\textrm{2.4}}(f)$, typically used for fitting spectral peaks. Its free parameters are the central frequency $f_0$, the width of the peak $\Delta f$ and the amplification factor $a_{2.4}$. Figure \ref{fig:spectral_fitting} shows the effect of each of the parameters.

\subsection{Spectral fitting algorithm}
The fitting is done using the displacement spectra of the P and S wave on the vertical component, using the pre-event noise, P and S wave windows defined by MQS.

We follow the following algorithm:
\begin{enumerate}
\item Determine the frequency range for which event data is clearly above pre-event noise, separately for P and S (\flow, \fhigh).
\item Fix $t^*$ such that the ratio of P- and S-wave is flat across all frequencies from $\flow$ to $\fhigh$, after correction for attenuation.

\item Fix $A_0$ such that the flat part of the S-wave spectrum below the 2.4 Hz resonance is matched.
\item Fix the S/P ratio $\SPratio$ such that the energy of the long-period part of the spectrum matches between P- and S-phase.

\item Fix \fc{} such that the decaying part of the S-wave spectrum (neglecting the 2.4 Hz resonance) is matched.
\item Fix $f_0, a_{2.4}, \Delta f$ such that the 2.4 Hz resonance of the S-spectrum is matched.
\item Iterate all above steps in order again until a match of both phase spectra is obtained, in which the predicted waveforms does not deviate by more than 5 dB from the observed one along the whole spectrum between \flow, \fhigh. 
\item As a final step, all parameters are varied manually around the best value to obtain uncertainty estimates.
\end{enumerate}
To perform the fitting, we use a self-developed graphical user interface available from github \cite{ceylan_marsspectgui_2025}.

\clearpage

\subsection{Fitted data set}
\begin{figure}[h]
    \centering
    \includegraphics[width=0.95\textwidth]{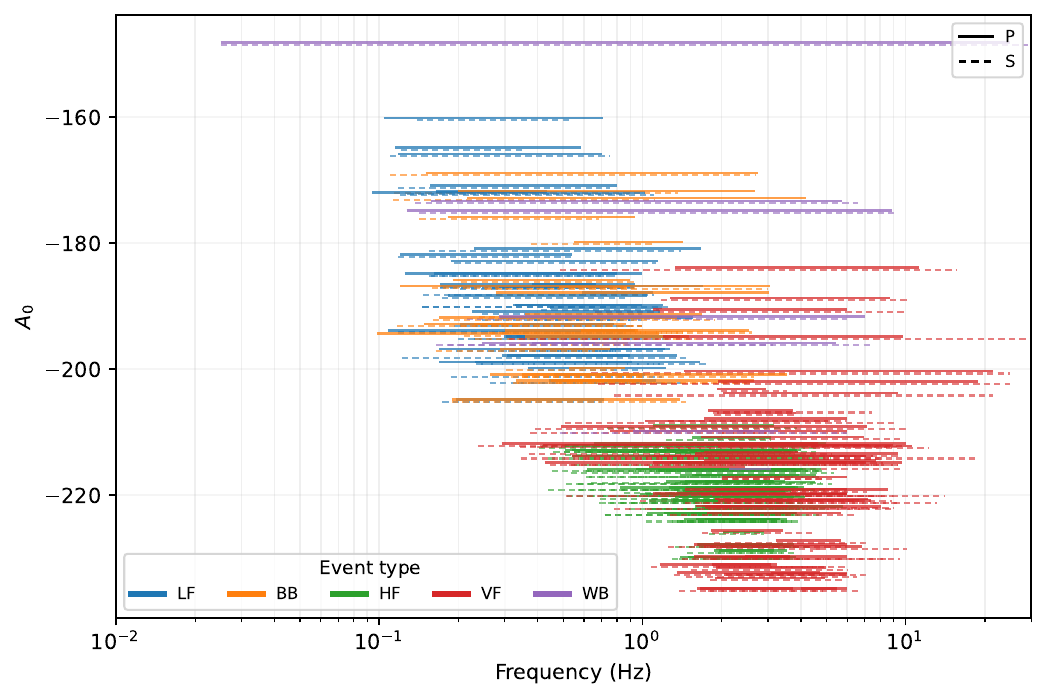}
    \caption{Frequency limits manually obtained for all events in this study, separately for P and S-waves (solid and dashed respectively).}
    \label{fig:freqlims}
\end{figure}
\begin{figure}[h]
    \centering
    \includegraphics[width=0.95\textwidth]{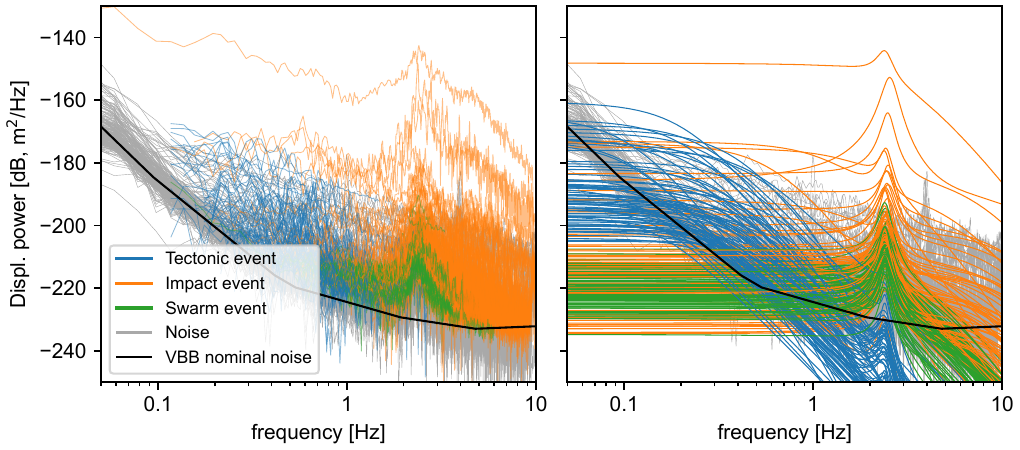}
    \caption{Observed S-wave spectra for all events in this study (left) are compared with model-based reconstructed spectra (right). Event types are color-coded: blue for LF and BB (later interpreted as tectonic), orange for WB and VF (impacts), and green for HF and 2.4 Hz (swarm quakes). The black line marks the nominal minimum noise of the VBB seismometer; grey curves show observed noise during quakes.  
 Tectonic quakes, with their low corner frequency, compete with rising low-frequency instrument noise. Hence, as compared with swarm quakes or impacts, in order to be detected, tectonic quakes require much higher long-period amplitudes—and thus magnitudes.}
    \label{fig:data_vs_fit}
\end{figure}

The v14 catalogue contains over 1300 events, a number that will increase to over 1900 events in the final catalogue in preparation, and manual reviewing of all events is not feasible. 
The majority of these events have spectra which are just barely above the background noise, and in particular the spectra of weak HF events only exceed the noise spectra in the narrow 2.4~Hz bandwidth. In these cases, we cannot fit the spectral parameters in a meaningful way. 
Therefore, we focus on high-quality events (Quality A\&B, termed QA and QB henceforth) from the MQS catalog while also including a representative set of lower-quality events (QC), but exclude poorly recorded events (QD). Using the above algorithm, we review all i) QA, QB \& QC LF and BB events  (81+2 WB); ii) all QA, QB \& QC VF events (63+7 WB); and iii) all QB HF events (74), as well as selected QC HF (8) and QB \& QC 2.4~Hz events (17) that are representative of the HF distance and magnitude range. Figure  \ref{fig:data_vs_fit} presents the observed and modelled spectra for all events that have been processed.

Although we determine the spectral parameters in a consistent manner, it should be highlighted that there are trade-offs among the free parameters and their uncertainties. Figure \ref{fig:freqlims} shows the frequency limits (\flow, \fhigh) obtained for all events. Most events are above noise for 2 octaves (factor 4 in frequency) or less, limiting the resolution of all seven parameters. In addition, several VF events have complex spectra that are not well-fitted with our simplified model. This will be discussed in Sect. \ref{sect:impact_peak}.

\section{Results}
\begin{figure}[h]
    \centering
    \includegraphics[width=0.5\textwidth]{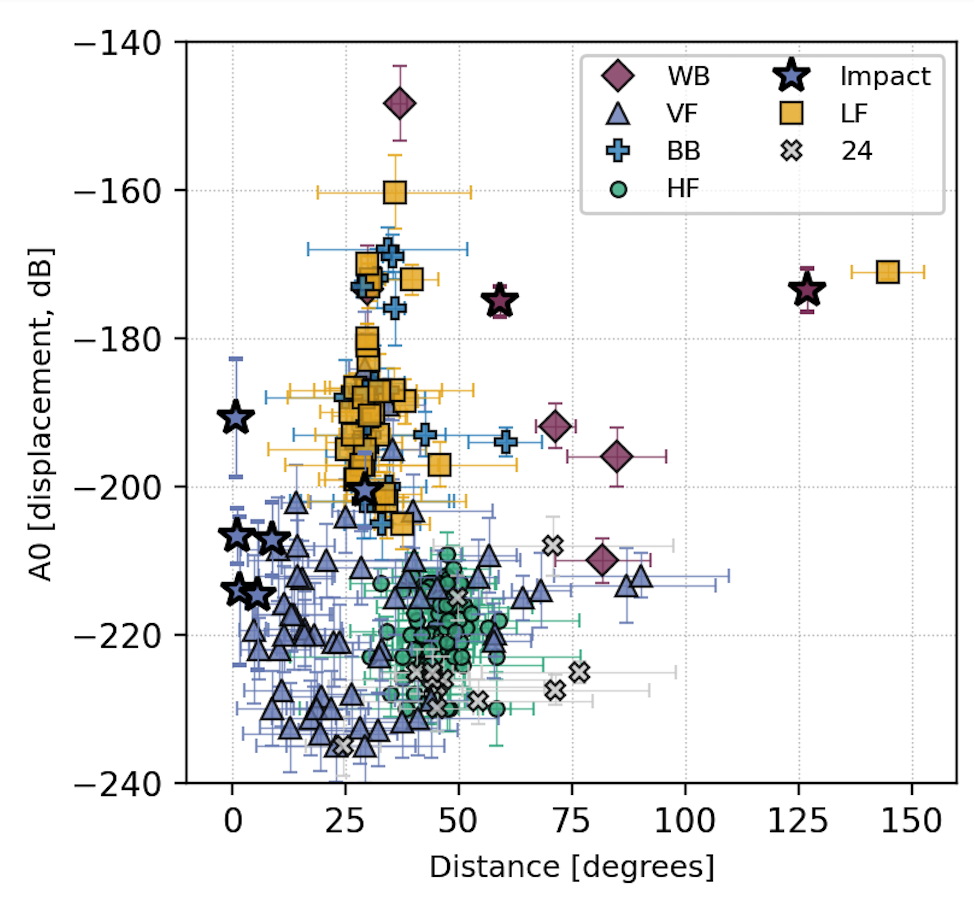}
    \caption{Distribution of long-period amplitude $A_0$ vs epicentral distance of the event. As evident, BB and LF events cluster between 28 and 34\textdegree distance, corresponding to the Cerberus Fossae graben system. VF and WB events are distributed over all distances, suggesting a non-tectonic origin, similar to the identified meteoritic impacts. The 2.4 Hz and HF events cluster between 40 and 55\textdegree{} distance, in a so-far unknown source region. }
    \label{fig:A0_distance}
\end{figure}

For the first time, this new fitting approach allows us to systematically determine a manually reviewed long period amplitude $A_0$, which should be robust against the effects of source spectral roll-off and attenuation encountered at higher frequencies, for all significant marsquakes recorded by InSight. As figure \ref{fig:A0_distance} demonstrates, the VF and HF events were mostly observed at lower amplitudes (consistent with figure \ref{fig:overview}) than LF or BB events. LF and BB events cluster tightly in distance around Cerberus Fossae (27.5 to 32.5\textdegree{}) \cite{clinton_marsquake_2021,stahler_tectonics_2022}. 

\subsection{Corner frequency}

\begin{figure}[h]
    \centering
    \includegraphics[width=\textwidth]{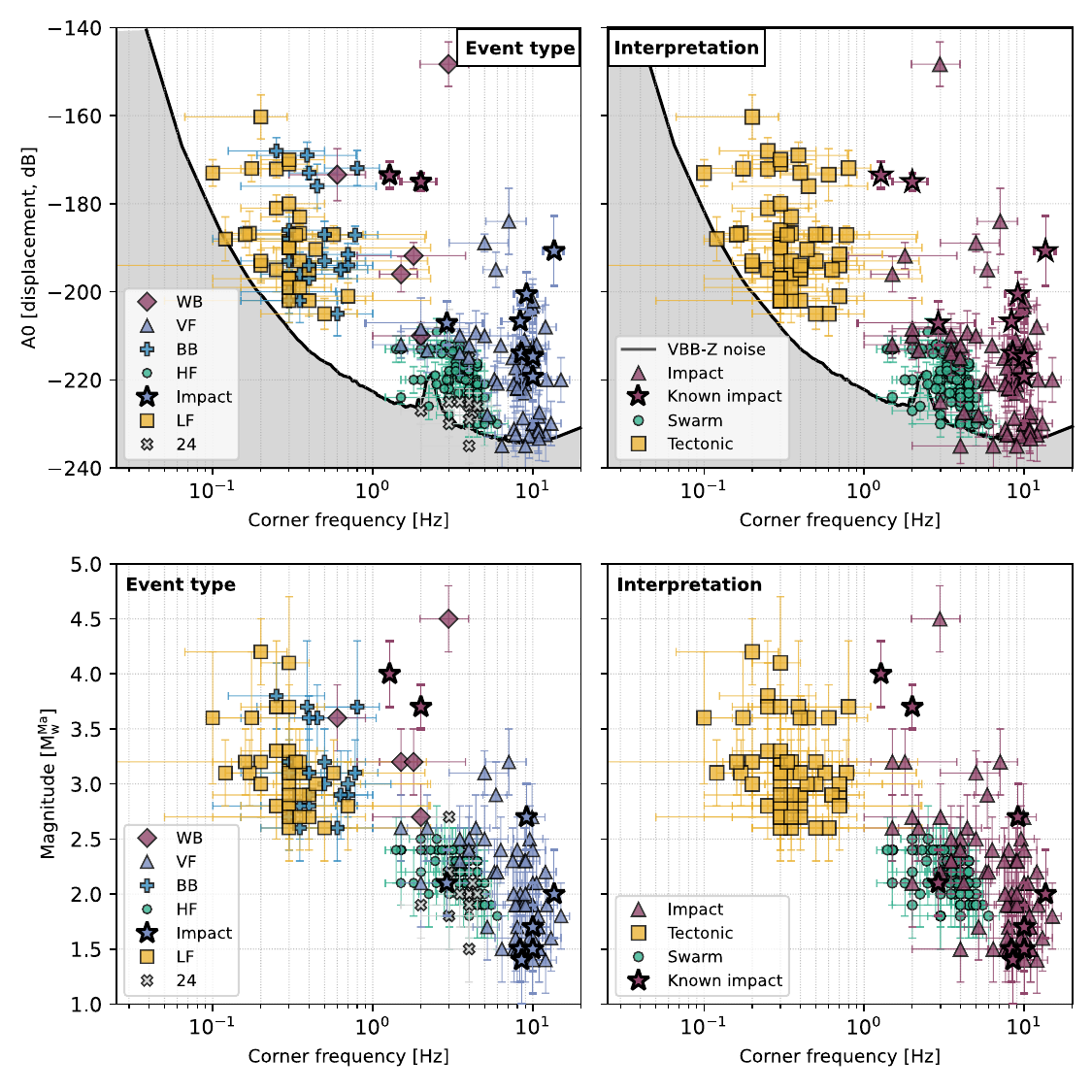}
    \caption{Distribution of corner frequency \fc{} vs. long period amplitude $A_0$ (top row) and magnitude (bottom row). The BB and HF marsquakes show a relatively low corner frequency, but follow a general trend of decreasing \fc{} with \MW. VF events and the confirmed impact-generated events are outside this relation. }
    \label{fig:clustering}
\end{figure}

The first investigation is into the relation between corner frequency \fc{} and the long-period energy plateau $A_0$  (see figure \ref{fig:clustering} a). The BB and LF marsquakes have corner frequencies below one Hz, with a slight trend of decreasing \fc{} with amplitude. Both event types overlap in this plot. HF events are a clearly separate group with significantly lower amplitudes, as well as corner frequencies between 1.5 and 5.5 Hz, with a trend of decreasing \fc{} with amplitude. The VF events span a much wider range of corner frequencies and amplitudes, with a slight overlap with the HF event group. A separate discussion is warranted for a class of BB events with relatively high corner frequencies, termed "Wideband" or WB in this figure. These contain the known big impacts, S1000a and S1094b, as well as other large, long-coda events, such as S1222a and S1102a. From their place in the A0/\fc{} distribution alone, they obviously fall outside the other event classes.

The long-period energy $A_0$ is dependent on both source strength and epicentral distance $\Delta$, so a more insightful parameter to look at is the moment magnitude \MW, which is a property only of the event. We use the spectral magnitude relation 
\begin{equation}
	M^{\textrm{spec}}_W = \frac{2}{3} \Bigl(\log_{10}A_0 + (1.0 \pm 0.1) \log_{10}\Delta + \left(12.6 \pm 0.5\right)\Bigr)
	\label{eq:MW_A0}
\end{equation}
defined in \citeA{bose_magnitude_2018, bose_magnitude_2021} to convert $A_0$ to moment magnitude (Fig. \ref{fig:clustering}), for an event at epicentral distance $\Delta$ (in degree). The scalar moment $M_0$ is then given by
\begin{eqnarray}
	M_0 &= & 10^{1.5 M_W + 9.1} \nonumber \\ 
	&=& A_0 \cdot \Delta \cdot 10^{21.7} \nonumber \\
	&\approx & A_0 \cdot \Delta \cdot (5 \cdot 10^{21})
	\label{eq:Moment_A0}
\end{eqnarray}
with $A_0$ in meter$/\sqrt{\textrm{Hz}}$ and $\Delta$ in degree (as usual in seismological magnitude formulae, the units do not match).

Given that the LF and BB events cluster in a tight distance range of less than 20\%, their distribution does not change significantly. For HF events, the same is true (distance range 28--60\textdegree{}), while the VF events, which cover all distances between 0.7 and 140\textdegree{}, now form a more distributed cluster.

Figure \ref{fig:clustering} c shows that the event types cluster strongly on the \fc{} vs \MW{} space. We will use this next to propose a new classification of event types. %
\begin{eqnarray}
\MW &\propto &2/3 \log_{10}(M_0)\\
&\propto &2/3\log_{10}\left(\fc^{-3}\right) \\
&\propto&-2\log_{10}\fc .
\end{eqnarray} 
 Given that the magnitude range of LF and HF events is only about one unit, not too much trust should be put into the best fitting $\MW= m \cdot \log_{10}\fc$ solutions. The VF events span across 4 magnitude units and a factor of 20 in \fc, so a clear relationship of decreasing \fc{} with magnitude is visible.

The WB events, i.e. the somewhat enigmatic class of events with high amplitudes and therefore moment magnitudes between 3 and 5., corner frequencies above 1 Hz, and extended coda, clearly do not fit into the cluster of LF and BB events. However, their corner frequencies follow the relation suggested by the VF events relatively well. Similar to the VF events, they have a long coda at high frequencies, indicative of a shallow source process. Two of these large events (S1000a and S1094b) have been confirmed as large meteoritic impacts \cite{posiolova_largest_2022}. The event S1102a has been tentatively located in a rugged area in Syrtis Major Planum \cite{zenhausern_lowfrequency_2022, ceylan_marsquake_2022}, where imaging conditions would not be beneficial to detect new craters. 

We therefore propose the following, simplified classification of marsquakes that takes into account the seismic evidence based on spectral fitting, but also the external information from tectonic studies and crater imaging.

\begin{enumerate}
    \item The only marsquakes that can certainly be interpreted as tectonic are the LF and BB events in Cerberus Fossae. For these events, there is a plausible tectonic environment \cite{perrin_geometry_2022}, based on the global stress distribution or the presence of a local mantle plume \cite{broquet_geophysical_2022}. 
    Additionally, a small number of marsquakes with short coda (S0183a, S0185a, S0976a) are plausible, somewhat deeper tectonic events. Note that this is a reduced list compared to \citeA{ceylan_mapping_2023}.
    For some of these events, it was possible to determine focal mechanisms based on P and S waveforms \cite{brinkman_first_2021, jacob_seismic_2022}.
    \item The largest group of marsquakes are the HF and 2.4 Hz events. Based on our model, there is no reason to separate the two. The distinction between the groups is a historical one and can be fully explained by the magnitude of the event and the signal-to-noise ratio. Note that this has always been the interpretation of the Marsquake Service \cite{van_driel_high-frequency_2021, clinton_marsquake_2021, ceylan_marsquake_2022}, although it must be admitted that this message was not always understood.
    As stated above, the interpretation that these events occur in the shallow region of Cerberus Fossae cannot be upheld. \citeA{dahmen_analysis_nodate} discusses their characteristics in more detail, but several of their features point to an as-of-yet unknown mechanism that is not a classical tectonic one. The events show a seasonality.
    \item Meteoritic impacts. As proposed by \citeA{zenhausern_estimate_2024}, VF events are plausibly caused by meteoritic impacts. Our more detailed analysis shows that the large broadband events (termed Wideband/WB above) belong rather to a class with the VFs than with other BBs or LFs.
\end{enumerate}

The right column of figure \ref{fig:clustering} illustrates the relationship between corner frequency and moment distribution using our newly developed classification scheme, which categorizes events into tectonic, swarm and impact types. This classification scheme provides a more refined approach for accurate event categorization and analysis, and will be adopted by the MQS for future implementations and studies.

\subsection{Seismicity}
\begin{figure}
    \centering
    \includegraphics[width=0.95\linewidth]{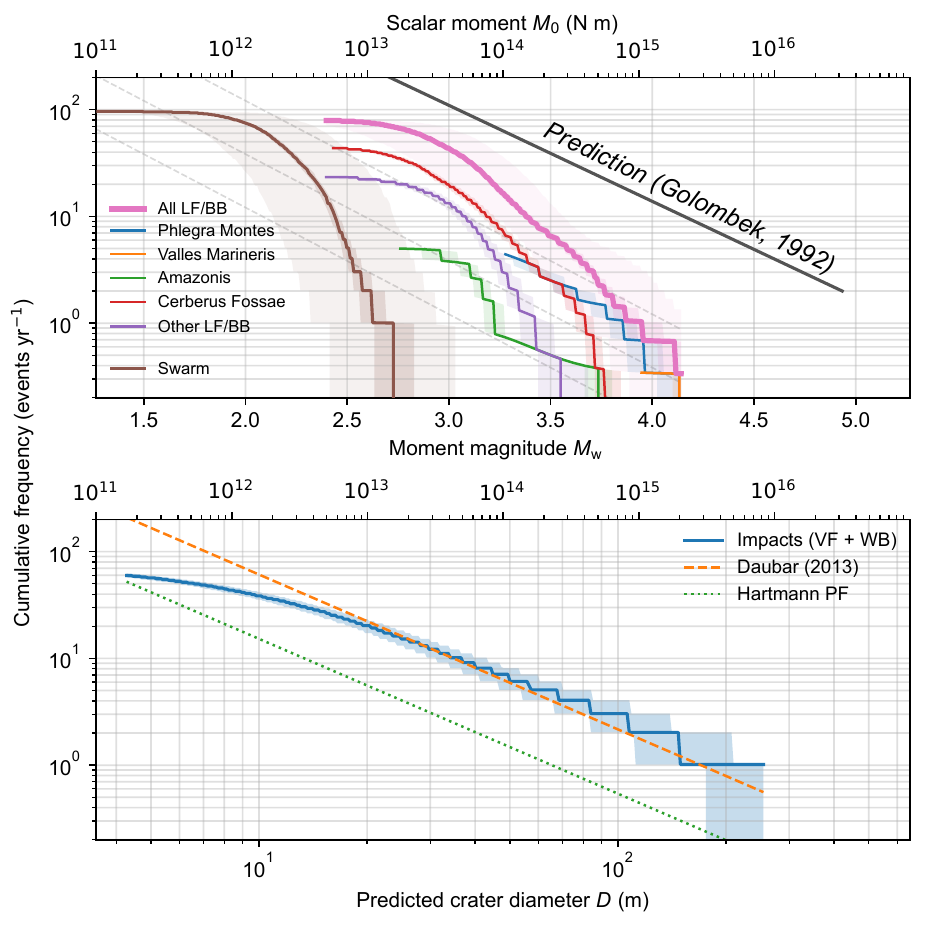}
    \caption{The seismicity of Mars: Top subplot: Size-frequency distribution of tectonic and swarm marsquakes per Earth year. The tectonic quakes are plot separately for the region of origin. Cerberus Fossae has the largest record of quakes, but a cutoff at magnitude 3.5. For Phlegra Montes, a small number of quakes above magnitude 3.5 have been observed on a single day \cite{nikolaj_louis_dahmen_seismicity_2024}, while no evidence of smaller events exists.
    Bottom subplot: The size-frequency distribution of meteoritic impacts follows the distribution proposed by \cite{daubar_current_2013} based on imaging.}
    \label{fig:seismicity}
\end{figure}

Using our new magnitude estimates from A0, we can constrain Martian seismicity by analyzing different event families separately. We update all event magnitudes based on the $A_0$ long period amplitude values determined above. For each event, we obtain an individual magnitude uncertainty estimate based on the uncertainty in $A_0$, and the uncertainty in distance $\Delta$. 
We use the estimate derived in eq. 5b in \cite{bose_magnitude_2021}:
\begin{eqnarray}
 \sigma_{M} =  0.44\,\sigma_{\log_{10}(A_0)}^2 + 0.044\left(\log_{10}\Delta\right)^2 + 0.44\,\sigma_{\log_{10}(\Delta)}^2 + 0.13  
\end{eqnarray}
Additionally, there is a correlated uncertainty in magnitude of 0.13, due to modelling uncertainty in derivation of our magnitude formulas \cite{bose_magnitude_2018, bose_magnitude_2021}. To estimate the overall seismicity, we create cumulative size-frequency curves, implementing the uncertainty in a Monte-Carlo manner: We create 10,000 individual curves, each perturbing the observed magnitude $M_{obs, i}$ into $M_{pert, i} = M_{obs, i} + \mathcal{N}(0, \sigma_{M,i})$, where $\sigma_{M,i}$ is the estimated uncertainty of event $i$. We finally perturb the whole set $\{M_{pert}\}$ by the correlated uncertainty of 0.13. From these 10,000 magnitude-frequency distributions, we then calculate the median curve, as well as 5th and 90th percentile for each magnitude bin. The result is plotted in figure \ref{fig:seismicity}.

We first consider tectonic (previously LF and BB) events originating from Cerberus Fossae (CF), which we define as all tectonic events with a back-azimuth pointing to CF, supplemented by all other tectonic events occurring at epicentral distances between 20° and 35°. A second group consists of the Phlegra Montes events, all recorded on Sol 1157. Although this cluster contains far fewer events, the occurrence rate of events with magnitudes of 3.5 and above even exceeds that inferred for Cerberus Fossae.

Finally, we use crater size estimates to constrain impact-related VF events. The confirmed impacts alone exhibit a production rate comparable to that predicted by the Hartmann/Daubar impact flux model; however, after including all the events we interpret as impacts, we infer a substantially higher overall impact-driven seismicity rate on Mars.
\clearpage

\subsection{S/P ratio}
\begin{figure}[h]
    \centering
    \includegraphics[width=0.65\textwidth]{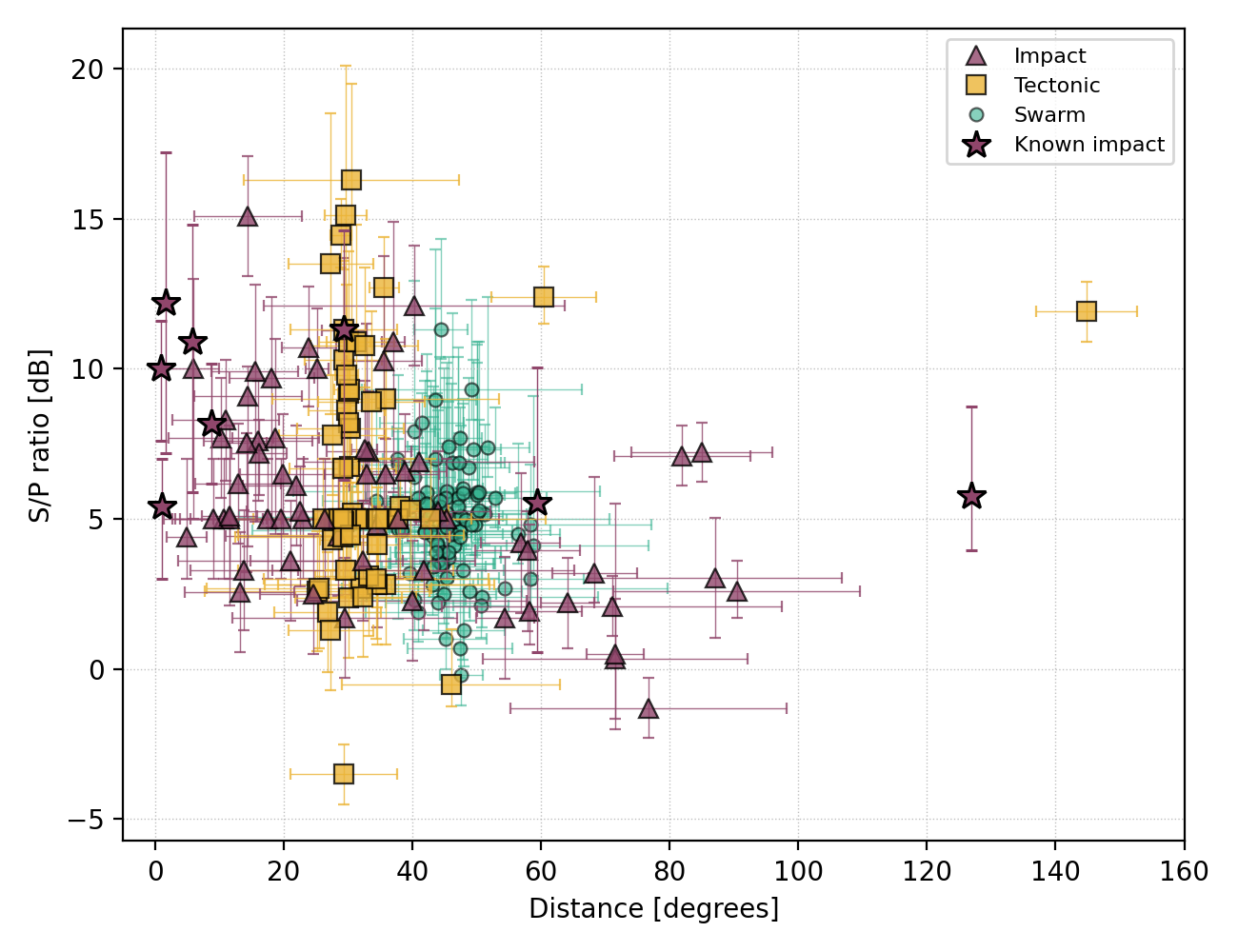}
    \caption{Distribution of S/P energy ratio in dB for all events, plot over distance. While some tectonic events show a distribution tending towards larger S/P ratios, there is no clear difference to the impact and swarm event clusters.}
    \label{fig:spectral_fitting_SP}
\end{figure}

Another parameter to investigate is the distribution of the S/P ratio $1/\SPratio$ (see figure \ref{fig:spectral_fitting_SP}). It shows that the S/P power ratio is on average 10 dB with a wide spread for tectonic events and 5 dB for swarm events. Impacts have a distribution similar to swarm events. The known small impacts near to InSight have a surprisingly high ratio, while the 2 confirmed distant impacts have some of the lowest ratios, more in-line with what would be expected for an impact source. 
If one assumes that the amplitudes of the tectonic events are least affected by propagation terms due to their clear mantle propagation, the S/P ratios can be compared to an expected term for a source in a Poisson medium (where the Lam\'e parameters are equal $\lambda=\mu$). For this, the power ratio should be 
\begin{equation}
\SPratio = \left(\frac{\alpha^3}{\beta^3}\right)^2 = 3^3= 27 \approx 15~\textrm{dB},
\end{equation}
when averaging over all source orientations and back-azimuths. Instead, we find $\SPratio\approx10$~dB. If this would be only explained by the source medium, it would require a ratio of P- to S-wave speed $\alpha/\beta\approx 1.3$. This value is not plausible, when compared to crustal values of 1.5-1.9 typically found on Earth. However, given that all significant tectonic events happened in a single source region, Cerberus Fossae \cite{stahler_tectonics_2022}, it is plausible that the strike direction is similar for all and that the low relative S-wave amplitude is an effect of source orientation. 

The swarm events show S/P ratios averaging to 5 dB with a spread between 1 and 10 dB. Given the lack of waveform similarity between the swarm events \cite{dahmen_analysis_2026}, it cannot be assumed that all have identical source mechanisms, so the argument made above for the low S/P ratio of the tectonic events does not hold here. If the Swarm events were a set of double-couple events with randomly distributed orientations, the S/P ratios should be distributed around 15 dB. Compensated Linear Vector Dipole (CLVD) source mechanisms, such as found during injection of fluids into a medium show relatively low S/P ratios \cite{walter_empirical_2007}. This is another indication that the HF swarm is not caused by classic tectonic earthquake processes.  

The confirmed and candidate meteoritic impacts show a slightly narrower distribution of S/P ratios than the tectonic events, with a few even reaching values around 0 dB or below. A simplistic model of a meteorite impact should create significantly stronger P-wave than S-wave energy and it was proposed to use this as a discriminator for impacts in the seismic catalog \cite{daubar_impact-seismic_2018}. However, within the uncertainty of the method, even several known impacts (marked by stars in fig. \ref{fig:spectral_fitting_SP}) show significantly higher S/P ratios than expected, of values of 10 dB or more. This suggests that P- to S-conversion near the source or effects of oblique impacts and resulting horizontal force transfer cannot be neglected and S/P ratio alone is not a certain discriminator of meteoritic impacts.

\clearpage
\subsection{Source scaling}
\begin{figure}[h]
    \centering
    \includegraphics[width=0.6\textwidth]{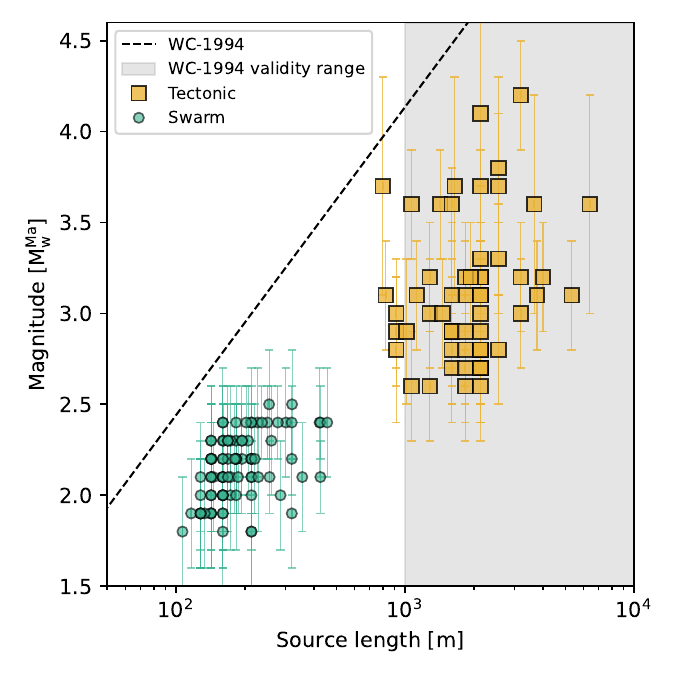}
    \caption{Distribution of source length from corner frequency for tectonic and swarm marsquakes, assuming a shear wave velocity $\beta=2$km/s near the source. The black line shows the WC-1994 scaling relation determined from aftershock patterns of 167 earthquakes \cite{wells_new_1994}. This dataset has been limited to rupture lengths $>1$km, marked by the grey area.}
    \label{fig:mag_vs_length}
\end{figure}

The spectral parameters can be used to obtain an estimate of the source size. The most straightforward conversion is the effective source size proposed by \citeA{brune_tectonic_1970}, 
\begin{equation}
    L=k \frac{\beta}{\fc{}}
\end{equation} where $L$ is the characteristic length of the fault, $\beta$ the shear-wave velocity, and $k$ a parameter that depends on the source geometry and the rupture propagation model. For circular sources, and \fc{} determined form S-wave spectra,\cite{brune_tectonic_1970} determined $k=2.34$. Using this factor, we arrive at source radii of 1 - 10 km for LF/BB events, 100-400 m for HF events, and 50-200 m for impacts. Figure \ref{fig:mag_vs_length} shows a comparison of these values with a magnitude-length scaling relation found from aftershock patterns on Earth \cite{wells_new_1994}. We generally find that our estimated source radii are a factor of 5-10 too large for a quake of a given magnitude. Given that we determined $L$ from corner frequency \fc, this is consistent with the low corner frequencies described above and previous work \cite{stahler_tectonics_2022}.

The mapping of surface fault segments in Cerberus Fossae from orbital images determined that most individual fault segments were shorter than 1 km \cite{perrin_geometry_2022}. It is implausible that large tectonic marsquakes would rupture multiple segments, which - once more - hints at a not-well-understood mechanism of low-frequency marsquakes.

\clearpage
\clearpage
\subsection{Crustal attenuation}
\begin{figure}[h]
    \centering
    \includegraphics[width=0.45\textwidth]{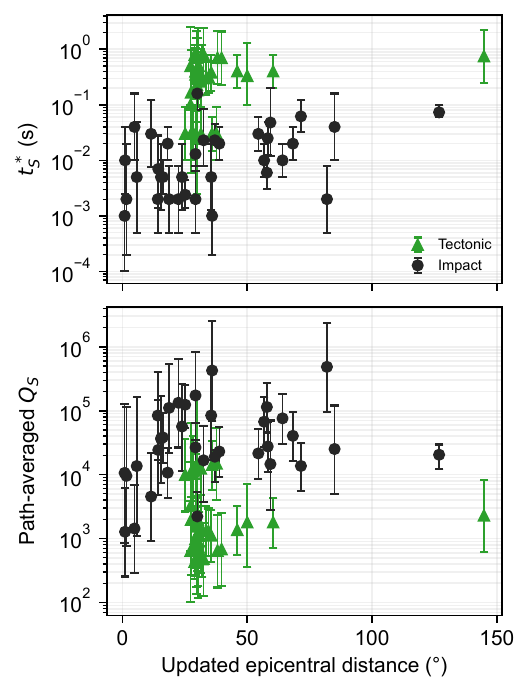}
    \caption{Path-averaged shear wave Q estimated for all tectonic and impact events. Note that this value is estimated only from the S-wave travel time and $t^*$, estimated in the fitting process. Given the highly scattered wavefield, the exact region of this \QS{} cannot be determined.
        }
    \label{fig:Q}
\end{figure}
Our estimates of attenuation, expressed by S-wave $t^*$, allow us to do a first quantification of attenuation in the Martian mantle. We calculate path-averaged shear-wave quality factors for the quality-A and quality-B impact events from the fitted attenuation parameter $t^*$ and the travel time. We use the asymmetric manually determined bounds $t^*_{\mathrm{low}}$ and $t^*_{\mathrm{high}}$ to obtain conservative uncertainty intervals on $\QS$. The resulting estimates indicate very weak attenuation for impact-related events: events at epicentral distances of up to approximately $40^\circ$ yield $\QS$ values of $10^4$ or higher. Beyond 40\textdegree distance, the estimated \QS{} values saturate. In many cases, the observed spectra constrain primarily a lower limit on $\QS$, rather than a finite preferred value. Our fitting method therefore does not robustly resolve the upper bound of the quality factor for these events, and the nearby observations are consistent with effectively negligible intrinsic attenuation over the sampled propagation paths.

\subsection{Crater sizes}
Our study allows us to present an updated table of crater size estimates from seismically detected meteoritic impacts. We follow the approach of \citeA{wojcicka_seismic_2020} and \citeA{ zenhausern_estimate_2024} to determine the crater rim diameter $D$ from the seismic moment $M_0$ as 
\begin{equation}
	D = \left(
	\frac{M_0}{(8.1\pm 3) \cdot 10^8}
	\right)^{1/3.3}.
\end{equation}
Combined with our magnitude formula (eq. \ref{eq:Moment_A0}), this yields a relationshop between long period displacement energy $A_0$ in m$^2$/Hz, distance $\Delta$ in degree and crater diameter $D$ in meter as
\begin{eqnarray*}
	D &\approx& \left(A_0\cdot \Delta\cdot 6.2\cdot 10^{13} \right) ^{1/3.3} \\
	 &\approx& 7.5 \cdot 10^{3} \left(A_0 \cdot \Delta \right) ^{0.3}.
\end{eqnarray*}
The uncertainty of this estimate is 
	 \begin{equation}
	 	\left|\frac{\delta D}{ D}\right| = 0.13 + 0.3  \left|\frac{\delta A_0}{ A_0}\right|,
	 \end{equation} 
where $\delta A_0/A$ is the relative uncertainty in determining the amplitude $A_0$. As discussed in \citeA{bose_magnitude_2021}, the distance error $\delta\Delta$ has a negligible effect on the magnitude- and hence crater diameter estimation. Table \ref{tab:crater_sizes} in the appendix shows updated size ranges for all events identified as impact candidates. 
 
\begin{figure}[h]
    \centering
    \includegraphics[width=0.75\textwidth]{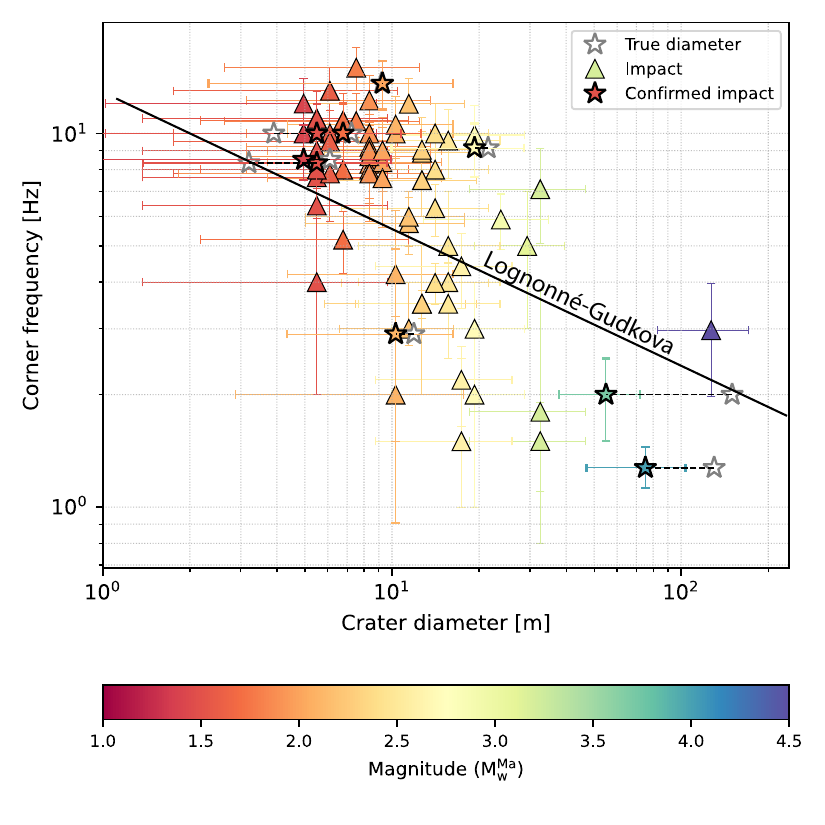}
    \caption{Estimate of crater diameter vs. corner frequency of all seismic events interpreted as impacts. The conversion from scalar moment $M_0$ to diameter $D$ uses the law $M_0=8.1\cdot 10^8 \times D^{3.3}$ established in \citeA{zenhausern_estimate_2024} from all confirmed seismic impact events.
    	The black line marks the relationship proposed in \citeA{charalambous_new_2025} building on the Gudkova-Lognonne source model for meteoritic impacts \cite{daubar_impact-seismic_2018, gudkova_impact_2015}.
        }
    \label{fig:crater_vs_fc}
\end{figure}

Figure \ref{fig:crater_vs_fc} shows the distribution of crater size estimates against measured seismic corner frequencies. Building on the source spectra model of \citeA{gudkova_impact_2015}, a recent crater detection paper \cite{charalambous_new_2025} presented a semi-empirical model of source duration $\tau$, depending on scaled impact momentum $I/I_0$: $\tau = 0.21 (I/I_0)^{0.14}$. Calibrated against several smaller impacts recorded by InSight and against the Saturn IV-B impacts recorded by the Apollo seismic experiment to determine impact momentum from crater size, they predict a corner frequency of 5.5 Hz corresponding to a 10 m crater on Mars and 2.3 Hz to a 100 m crater. 
	Note that we cannot fully reproduce the authors' determinations of source duration and corner frequency for several nearby impacts, which also come with surprisingly small uncertainties. We generally find that our (relatively large) uncertainties in corner frequency envelop the predicted values from the model.

\subsection{Spectral peak for impact candidates}
\label{sect:impact_peak}
\begin{figure}
    \centering
    \includegraphics[width=0.95\linewidth]{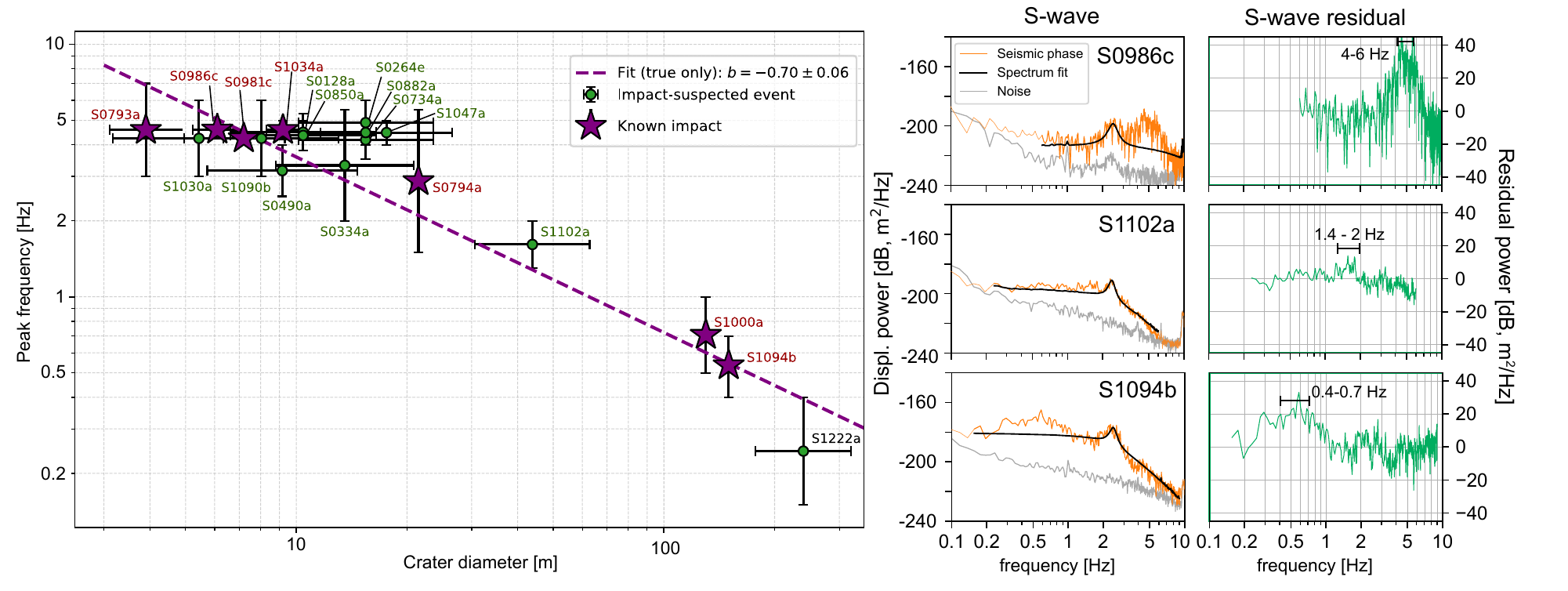}
    \caption{Left: Frequency of the additional spectral peak observed for several suspected impact events vs. observed (star) or inferred (circle) crater diameters. Right: Examples of fits for three events: Known near-by impact S0986c (6 meter, 1.2\textdegree{} distance), suspected impact S1102a (est. 44 meter, 73\textdegree), known large impact S1094b (150 meter, 60\textdegree). }
    \label{fig:spectral_peak}
\end{figure}
Fitting the spectra of Marsquake signals to identify meteoritic impacts involves a detailed analysis of the observed displacement spectrum, particularly focusing on the energy distribution within the 2 to 6 Hz frequency range, while considering the entire range of 0.5 to 10 Hz. The meteorite impact candidates, as delineated in our previous model, provide a robust framework for this spectral fitting. Notably, confirmed nearby impacts consistently exhibit an overshoot in the displacement spectrum, characterized by a surplus of energy between 3 and 7 Hz (e.g. S0986c, see fig. \ref{fig:spectral_peak}B), deviating from the predictions of the Brune model. This spectral anomaly is less pronounced but still discernible in distant large impacts, such as the event labeled S1094a, where it occurs around 0.5 Hz (see fig. \ref{fig:spectral_peak}C). Intriguingly, several other events that have not yet been conclusively identified as impacts also display this overshoot feature. This "overshoot" has been discussed in \citeA{daubar_impact-seismic_2018}, in the SWH model, without specifying a mechanism.

\begin{figure}
    \centering
    \includegraphics[width=0.9\linewidth]{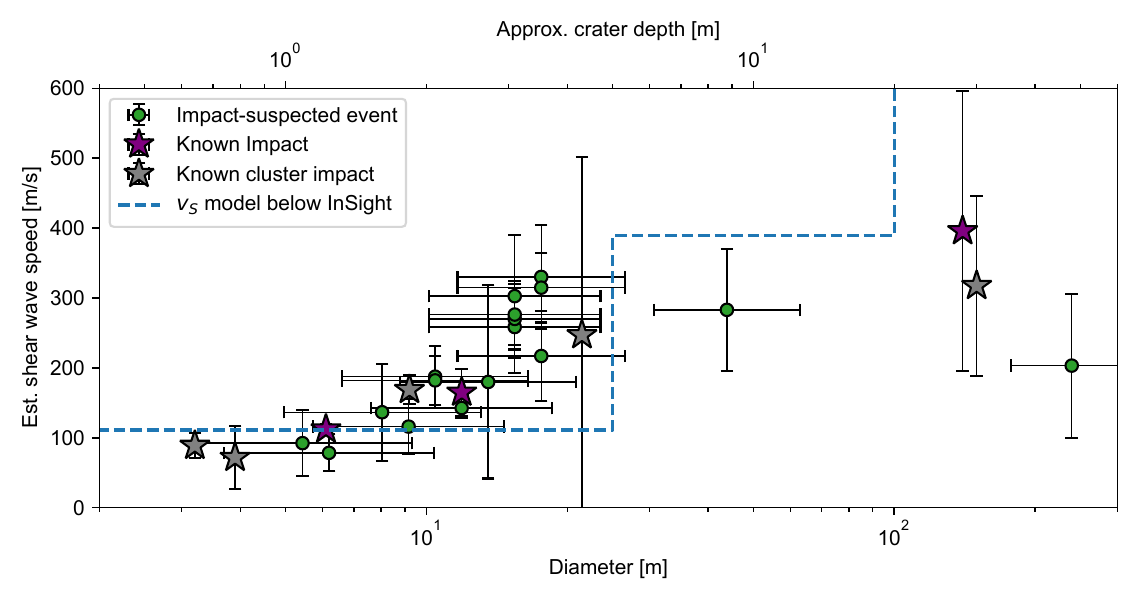}
    \caption{Crater diameter $D$ and depth $h$ vs estimated shear wave speed based on frequency $f_0$ of spectral "bump", $\vS = 4 D f_0$. The dashed line marks the near surface \vS{}-model of \cite{hobiger_shallow_2021} determined for the InSight landing site.}
    \label{fig:crater_depth}
\end{figure}

A simple way to rationalize the additional spectral peak is to interpret it as the fundamental quarter-wavelength resonance of an impact-generated compliant (low-\vS) near-source damage zone of effective thickness $H$. In a 1-D approximation, the fundamental frequency of this resonance is $f_0 \approx \vS/(4H)$. If the damage thickness scales with the diameter of the crater $D$ as $H=\alpha D$, then 
\begin{equation}
    f_0 \approx \vS /(4\alpha D)  \propto 1/D.
\end{equation}
If one fixes $\alpha = 1$, i.e. assumes that the damaged zone is similar in depth to the crater diameter, one can estimate the shear wave speed at the impact site as $\vS = 4 D f_0$. Table \ref{tab:peak_vs_cratersize} shows estimated values for the peak frequency, as well as the derived shear wave speeds. We find that the values generally scatter around 250 -- 300 meter per second, with a few outliers to significantly lower values. In general, larger craters (S0794a, S1000a, S1094b) favour higher shear wave speeds, whereas smaller craters (S0793a, S0986c) show much lower estimates of only 70 -- 100 m/s. If we convert crater diameter into depth with a conversion factor of $h=0.2D$, as proposed by various authors \cite{millot_depth_2025} , we find an increase in estimated \vS{} at crater depths of about 3 meters (see fig \ref{fig:crater_depth}). It is generally plausible that smaller events interact only with weak materials near the surface, with lower shear moduli and hence \vS{}, as reported for the uppermost few meters of the InSight landing site \cite{hobiger_shallow_2021, brinkman_situ_2022, delage_investigating_2023}.

While a general quantitative match can be made, the exact mechanism of this peak remains speculative. More recent modeling efforts of meteoroid impact spectra \cite{froment_numerical_2024} did not treat material deformation specifically. 
Near-source damage and spall layers generated by explosions have been shown to strongly modify seismic spectra through reverberation and interference within a compliant near-surface zone \cite[termed spall layers there]{day_seismic_1991}. Such a structure is expected to support resonant behavior at frequencies controlled by its thickness and elastic properties. 
\begin{table}[]
    \centering
    \begin{tabular}{c|rlll}
    Event name & Distance & Crater diameter & Peak frequency & \vS{} [m/s]  \\
    &  &  [m] & [Hz] & estimate \\
S0128a (B) & 7.8\textdegree & 10 (est.)& $4.5 \pm 1 $ & $ 187 \pm  44$\\ 
S0202b (C) & 14.1\textdegree & 6 (est.)& $3.2 \pm 1 $ & $  78 \pm  26$\\ 
S0218a (B) & 31.1\textdegree & 12 (est.)& $3 \pm 0.3 $ & $ 142 \pm  13$\\ 
S0264e (B) & 34.5\textdegree & 15 (est.)& $4.9 \pm 1 $ & $ 303 \pm  88$\\ 
S0334a (B) & 19.8\textdegree & 14 (est.)& $3.3 \pm 3 $ & $ 180 \pm 138$\\ 
S0421a (B) & 36.8\textdegree & 18 (est.)& $4.7 \pm 1 $ & $ 330 \pm  75$\\ 
S0424c (B) & 31.2\textdegree & 18 (est.)& $3.1 \pm 0.9 $ & $ 217 \pm  65$\\ 
S0490a (B) & 17.4\textdegree & 9 (est.)& $3.2 \pm 1 $ & $ 116 \pm  39$\\ 
S0533a (C) & 5.5\textdegree & 12 (meas.)& $3.5 \pm 0.7 $ & $ 165 \pm  34$\\ 
S0734a (B) & 8.3\textdegree & 15 (est.)& $4.2 \pm 1 $ & $ 258 \pm  66$\\ 
S0793a (B) & 3.0\textdegree & 4 (meas.)& $4.6 \pm 3 $ & $  71 \pm  45$\\ 
S0794a (B) & 17.6\textdegree & 22 (meas.)& $2.9 \pm 3 $ & $ 247 \pm 255$\\ 
S0850a (B) & 8.3\textdegree & 10 (est.)& $4.4 \pm 0.9 $ & $ 182 \pm  36$\\ 
S0864b (C) & 25.1\textdegree & 15 (est.)& $4.4 \pm 0.7 $ & $ 270 \pm  44$\\ 
S0882a (B) & 21.7\textdegree & 15 (est.)& $4.5 \pm 0.7 $ & $ 276 \pm  44$\\ 
S0986c (B) & 1.2\textdegree & 6 (meas.)& $4.6 \pm 0.3 $ & $ 112 \pm   7$\\ 
S1000a (A) & 128.3\textdegree & 140 (meas.)& $0.71 \pm 0.4 $ & $ 396 \pm 201$\\ 
S1030a (B) & 11.2\textdegree & 5 (est.)& $4.2 \pm 2 $ & $  92 \pm  47$\\ 
S1034a (B) & 0.8\textdegree & 9 (meas.)& $4.6 \pm 0.6 $ & $ 169 \pm  21$\\ 
S1047a (B) & 13.6\textdegree & 18 (est.)& $4.5 \pm 0.7 $ & $ 315 \pm  50$\\ 
S1090b (B) & 9.8\textdegree & 8 (est.)& $4.2 \pm 2 $ & $ 136 \pm  69$\\ 
S1094b (A) & 59.7\textdegree & 150 (meas.)& $0.53 \pm 0.2 $ & $ 317 \pm 129$\\ 
S1102a (A) & 73.3\textdegree & 44 (est.)& $1.6 \pm 0.5 $ & $ 283 \pm  87$\\ 
S1160a (C) & 2.1\textdegree & 3 (meas.)& $6.9 \pm 1 $ & $  89 \pm  18$\\ 
S1222a (A) & 37.0\textdegree & 239 (est.)& $0.21 \pm 0.1 $ & $ 203 \pm 103$\\ \\
    \end{tabular}
    \caption{Crater diameters measured by orbital imaging or estimated from seismic moment compared to the frequency of the additional spectral peak. The shear wave speed is estimated from assuming a simple $\lambda$/2 resonance.}
    \label{tab:peak_vs_cratersize}
\end{table}
This recurring pattern of enhanced spectral energy in the specified frequency range suggests it could serve as a diagnostic tool for distinguishing meteoritic impacts from other seismic events on Mars, enhancing our understanding of Mars's seismic activity and the impact history.

\section{Discussion}
\subsection{Seismicity}
The level of seismicity observed by InSight can now be constrained for the complete dataset spanning nearly 4 Earth years long. Compared to the first published estimate in \citeA{banerdt_initial_2020}, the rate of observed events is slightly increased: Globally, about 10 marsquakes of magnitude 3.5 or above were observed. Compared to the pre-InSight estimate of \citeA{golombek_prediction_1992}, we find about 5 times less events than predicted. \citeA{knapmeyer_working_2006} had presented 4 different models to predict the level of seismicity based on global fault mapping. These models were differentiated by the maximum magnitude (termed \textit{few faults} for a high maximum magnitude of about 6 and \textit{many faults} for a lower maximum magnitude of 3.5). Additionally, the authors estimated the effect of lithospheric strength on seismicity. Our observation of 50 events of magnitude 3 or above matches the model for a weak lithosphere and many faults, over which seismicity is distributed. Specifically for the Cerberus Fossae system, we find a significant lack of events > 3.5. Given our 20 M>3 events, we would have expected about 2 events at magnitude 4, of which we found none, even though we can expect to detect magnitude 4 events at all times of day and year. The magnitude/frequency distribution of Cerberus Fossae events (orange in fig. \ref{fig:seismicity}) thus shows a catalog that is complete above magnitude 3 and shows a roll off at magnitude 3.6.

\subsection{
Identification of VF Events as Meteoritic Impacts}

We interpret the majority of Very High Frequency (VF) events as seismic signals generated by meteoritic impacts. The strongest evidence is provided by the direct association of several smaller VF events with fresh impact craters identified in orbital images \cite{garcia_newly_2022, bickel_new_2025, charalambous_new_2025}. For these events, the crater locations are also consistent with independently inferred seismic distances and, where measurable, back azimuths. These confirmed cases demonstrate that impacts can produce observable seismic events, with specific characteristics, such as high corner frequencies compared to the event magnitude, and extended codas. 

The spectral properties of the broader VF population support the same interpretation. VF events span a wide range of amplitudes and epicentral distances, but in a \fc{} vs \MW plot, they form a distinct cluster, which envelopes all previously observed impact-seismic events. Additionally, VF events do not occupy the compact distance and spectral distributions of either the tectonic LF/BB population or the HF swarm. Their high corner frequencies indicate short source durations and shallow, spatially compact sources, while their long high-frequency codas are consistent with strong scattering in the fractured and porous material surrounding a surface impact. Numerical impact models likewise predict cutoff frequencies and seismic moments comparable to those observed for the confirmed VF impacts  \cite{rajsic_seismic_2023, froment_meteorite_2023}. It has also been shown that while the rate of VF events exceeds the annual number of fresh craters observed from orbital imaging \cite{zenhausern_estimate_2024}, the resulting rate of impacts is compatible with earlier estimates extrapolated from lunar cratering \cite{hartmann_cratering_2001}. While these estimates had been seen as an upper limit until recently, re-processing of images of the Mars Reconnaissance Orbiter context camera CTX using machine-learning based detection algorithms \cite{wagstaff_using_2022} led to a significant increase in new crater detections \cite{bickel_new_2025}. The normalization of the rate of impacts from an incomplete imaging (or seismic) dataset is entirely non-trivial and hence warrants further research.

A recent study \cite{garcia_detection_2025} found statistically significant waveform similarity between several small VF events (S0334b, S0334c, S0343a), which was interpreted as a tectonic nest of marsquakes. The waveform similarity observed by the authors can only be explained by an almost identical source location and process, which rules out a meteoritic impact source. In our analysis, we were not able to obtain a stable spectrum for S0334c, given the low SNR, while we identify S0334b, S0343a as meteoritic impacts in 35.9\textdegree and 37.7 \textdegree distance with crater sizes of 3-7 and 4-8 meters respectively. Within the uncertainty of the MQS distance estimates, the two events could as well be co-located. Both events have a low SNR and a general similarity to other VF events and would hence be classified as impacts in our scheme. \citeA{garcia_detection_2025} are entirely correct to point out that two events with similar source locations, waveform similarity and only nine days of separation are unlikely impact candidates. However, no other such effect has been found for any VF event pair, specifically not to one with a higher SNR, so we mark the observation as a call to treat our results with care.

No single seismic characteristic uniquely identifies an impact: high-frequency energy and long codas may also occur for shallow tectonic sources. Taken together, however, the direct crater associations, the broad geographic distribution of VF events, their shallow-source waveform character, and their continuous spectral scaling with independently confirmed impacts provide a parsimonious argument that most VF events share an impact origin. We therefore treat VF as an observational waveform class whose dominant physical source is meteoritic impact, while retaining the possibility that individual low-quality or spectrally ambiguous events may have another origin.

\subsection{S1222a}
An enigma remains the largest marsquake ever observed, S1222a \cite{kawamura_s1222alargest_2023}. It has strong surface waves \cite{kim_structure_2023, panning_locating_2023} and an extended coda similar to other observed impact-generated quakes. However, a preliminary review of orbital images did not yet find a fresh crater close to the seismically constrained epicenter \cite{fernando_tectonic_2023}. This search was focused on orbital detection of the impact blast zone, in effect searching for an area in which dust is removed, leading to a significantly reduced albedo. This is done because the blast zone can extend over a significantly larger area than the size of the crater itself. However, it is unclear whether an impact in a dust-free environment would even produce a significant visible blast zone. Only a good third of the potential epicentral region was additionally searched with images from the CTX camera onboard Mars Reconnaissance Orbiter \cite{malin_context_2007}, whose resolution would be high enough to detect the crater itself. Further, some investigations of surface wave polarization \cite{kim_structure_2023, maguire_focal_2023} pointed to a source region south of -10\textdegree S, i.e. outside the search area of \citeA{fernando_tectonic_2023}.
In the opinion of the authors, the seismic evidence points to a meteoritic impact as the source of the S1222a event, although this cannot be confirmed without the detection of the actual impact crater. With time passing, the crater - if it exists - will be covered by dust again, but for the next ten years, it should still stick out as a relatively fresh structure. Continuous, semi-automated review of CTX images as described in \citeA{bickel_new_2025} should be able to spot it eventually.

Our analysis of marsquake spectra extended to the full InSight dataset confirms the persistence of the previously observed horizontal-to-vertical energy ratio (H/V).  Studies \cite{carrasco_empirical_2023, hobiger_shallow_2021} had inverted this H/V curve for a layered structure below the receiver. They had found that a sandwich structure of several rigid vs soft layers is necessary, specifically to suppress H/V values $>1$ around the 2.4 Hz H/V notch. These layers were interpreted as volcanic and sedimentary layers respectively, stemming from the intermittent volcanism in the Elysium region since the Hesparian age. Our analysis confirms the dataset which their inversions were based on.

\subsection{Swarm events}
For the Swarm events, our study shows a corner frequency smaller than the expected value for events of that magnitude, resulting in relatively large source length estimates. The mechanism behind the swarm events remains enigmatic. \citeA{dahmen_analysis_2026} demonstrated a robust seismicity, as well as a two-fold increase in seismicity in the second Martian year, without providing a plausible explanation. \citeA{shi_near-surface_2025} recently proposed a mechanism related to near-surface water ice melting in Northern summer. While this could explain the seasonality of the events, it does not explain the source of the deformation to provide enough strain for repeated events with characteristic source lengths of 10--50 meter. 

\subsection{Attenuation}
Our attenuation results are broadly consistent with the high crustal $\QS{}$ inferred from the first InSight high-frequency events by \cite{lognonne_constraints_2020, menina_energy_2021, van_driel_high-frequency_2021}, who reported values exceeding approximately $5000$, based on a smaller, inital dataset of HF and VF events. Our analysis incorporates a larger dataset of impact events and explicitly accounts for asymmetric uncertainties in $t^*$, and tries to reduce the strong trade-off between source spectral decay and attenuation. 

Since the spectra were computed in an extended time window in the S-wave coda of up to 60 seconds length, it is important to remember that they cannot be directly mapped to a ray theoretical propagation path. Typical Mars mantle models \cite{duran_seismology_2022, khan_evidence_2023, samuel_geophysical_2023, drilleau_marsquake_2022} show relatively low velocity gradients with depth, hence scattered waves can have diverged significantly from the theoretical path of a first-arriving phase. Specifically, propagation of the coda can have occurred predominantly in a lower crustal or lithospheric layer with very low attenuation.
The observed VF events at distances above 40\textdegree{} require a \QS{} of at least 10000, which is at the upper end to crustal \QS{} values of 7000--15000 observed for the Moon around 8~Hz \cite{garcia_lunar_2019}.
\citeA{nakamura_seismic_1982} had found estimated a frequency dependence of lunar shear-wave attenuation in the form $\QS{} = Q_0 \times f^{0.7\pm0.1}$. Since our measurements are span the frequency range between 3 and 8 Hz, we can try to attend for a frequency dependency of $Q$, but find that our spectra are fit best with a constant value. 

Interesting enough, the tectonic events, which are generally interpreted to represent deeper sources \cite{duran_seismology_2022} show generally stronger effects of attenuation. However, given the relatively low \fc{}-values, it is possible that an unresolved tradeoff between our estimates of corner frequency and attenuation remains.

\section{Conclusion}

We present a consistent spectral characterization of the significant seismic events recorded during the full InSight mission. By separating source amplitude, source duration, attenuation, and local amplification, the resulting dataset provides a unified quantitative description of marsquakes and serves as a toolbox for future studies of the complete catalogue, that may focus on deviations of individual events from this simple model. We present an interpretation framework: Low-frequency and broadband events near Cerberus Fossae are the most plausible \textit{tectonic} marsquakes, while most VF events align with independently confirmed \textit{meteoritic impacts}. HF and 2.4 Hz events form a distinct \textit{swarm} population whose properties are difficult to reconcile with conventional shear faulting. These results demonstrate that Martian seismicity reflects multiple source mechanisms and cannot be described by a single terrestrial analogue or magnitude-frequency relation.

Across all populations, corner frequencies are generally lower than expected for comparable terrestrial events, implying relatively large source dimensions and low apparent stress drops. For tectonic events, this may indicate rupture in warm or mechanically weak crust, while for the swarm population it suggests a fundamentally different source process. 

Impact-generated signals further show that intrinsic attenuation in the Martian crust and lithosphere is extremely low. Many events constrain only lower bounds on ($Q_S$) exceeding $10^4$. Despite strong scattering, seismic energy persists to several Hertz over planetary distances, indicating that much of the observed spectral decay originates at the source rather than during propagation.

Separating event populations also refines estimates of Martian seismicity. Tectonic activity is dominated by Cerberus Fossae, which however lacked events larger than magnitude 3.5–3.7 over the three years of InSight, despite their detectability. Other regions produce fewer but occasionally larger events, such as Phlegras Montes, a region not interpreted as seismic before the mission. Overall, the number of large marsquakes is 2-5 times lower than many pre-InSight predictions. The inferred impact population broadly follows expected scaling, though uncertainties remain due to incomplete detection and crater identification.

Some interpretations remain tentative, including the origin of the HF swarm, the additional spectral peaks in impact candidates, and the classification of S1222a. Nonetheless, their consistent spectral behavior provides testable predictions for future observations and modeling.

InSight has brought single-station planetary seismology close to its practical limit. With only one station, key source and propagation parameters remain coupled, and some ambiguity is unavoidable. While further analysis will continue to refine the dataset, major advances will require a seismic network. Until then, the spectral catalogue presented here offers a comprehensive framework for interpreting the InSight record and guiding future exploration of Martian seismicity.

\section{Open Research}
The InSight event catalog V14 (comprising all events until the end of the mission) and waveform data are available from the Incorporated Research Institutions for Seismology Data Management Center (IRIS-DMC), National Aeronautics and Space Administration Planetary Data System (NASA PDS), SEIS-InSight data portal, Institut de Physique du Globe de Paris (IPGP) data center (\url{https://www.insight.ethz.ch/seismicity/catalog/v14}) and MarsQuake Service catalog by InSight Marsquake Service (2023). The software package \textit{MarsSpectGUI} used for fitting is available on github \cite{ceylan_marsspectgui_2025}.

\acknowledgments
The authors acknowledge the National Aeronautics and Space Administration (NASA), CNES, partner agencies and institutions (UKSA, SSO, DLR, JPL, IPGP-CNRS, ETHZ, ICL, MPS-MPG) and the operators of Jet Propulsion Laboratory. Finally, we acknowledge the contributions of all of our colleagues on the InSight mission. 

This paper is InSight Contribution Number 353.

\bibliography{library}

\appendix

\subsection{H/V ratios}

\begin{table}[hbt]
\centering
  \begin{tabular}{l|cc}
   f [Hz]& H/V ratio& event type\\ \hline
    0.3& 1.56 & LF\\
0.5& 1.36& LF\\
0.75& 1.22& LF\\
1.0& 1.13& LF \& HF\\
1.5& 1.22& HF\\
1.75& 1.17& HF \& VF\\
1.9& 1.11& HF \& VF\\
2.0& 1.01& HF \& VF\\
2.15& 0.846& HF \& VF\\
2.25& 0.749& HF \& VF\\
2.3& 0.731& HF \& VF\\
2.40& 0.762& HF \& VF\\
2.5& 0.882& HF \& VF\\
2.75& 1.08& HF \& VF\\
3.15& 1.20& HF \& VF\\
3.50& 1.39 & VF\\
3.8& 1.63 & VF\\
4.0& 1.70 & VF\\
4.35& 1.62 & VF\\
4.75& 1.84 & VF\\
5.25& 2.30 & VF\\
5.75& 2.65 & VF\\
6.4& 3.50 & VF\\
6.9& 4.62 & VF\\
7.5& 6.10 & VF\\
8& 7.11 & VF\\
9.16& 6.92 & VF\\
\end{tabular}
  \caption{H/V ratios for S-wave determined from all event types, as plotted in figure \ref{fig:HV_curve}.}
  \label{tab:HV}
\end{table}

\subsection{Crater size estimates}
\begin{longtable}{|l|clr|}
  \caption{Updated crater sizes based on estimates of scalar moment $M_0$. Two events have been identified as a tectonic nest by Garcia et al. (2025) and are marked with dagger. } \\
  \label{tab:crater_sizes}\\
  \hline
 Event name & Distance & $M_0 / 10^{12}$ Nm & Est. Diameter \\
 \hline
 \endfirsthead
	
 \hline
 \multicolumn{4}{|c|}{Continuation of Table \ref{tab:crater_sizes}} \\
 \hline
 Event name & Distance & $M_0$ / $10^{12}$ Nm & Est. Diameter \\
 \hline
 \endhead
 
 \hline
 \endfoot
 
 \hline
 \hline\hline
 \endlastfoot
S0128a & 14.4\textdegree & $  2.5 \pm   1.8$ & $    7-   13$ m\\
S0202b & 26.4\textdegree & $  0.6 \pm   0.3$ & $    4-    8$ m\\
S0218a & 57.9\textdegree & $  3.1 \pm   1.6$ & $    9-   15$ m\\
S0226a & 19.6\textdegree & $  0.2 \pm   0.1$ & $    3-    7$ m\\
S0234b & 31.6\textdegree & $  0.3 \pm   0.1$ & $    3-    7$ m\\
S0241a & 29.5\textdegree & $  0.3 \pm   0.2$ & $    3-    7$ m\\
S0263a & 11.5\textdegree & $  0.7 \pm   0.4$ & $    5-    9$ m\\
S0264e & 64.1\textdegree & $  6.0 \pm   2.0$ & $   11-   19$ m\\
S0334a& 35.9\textdegree & $  3.7 \pm   1.9$ & $   10-   17$ m\\
S0334b\dag & 32.3\textdegree & $  0.4 \pm   0.2$ & $    3-    7$ m\\
S0343a\dag & 37.7\textdegree & $  0.6 \pm   0.3$ & $    4-    8$ m\\
S0376a &  5.9\textdegree & $  0.3 \pm   0.1$ & $    3-    7$ m\\
S0387a & 58.2\textdegree & $  3.1 \pm   1.0$ & $    9-   15$ m\\
S0401b & 72.9\textdegree & $  2.0 \pm   0.3$ & $    9-   14$ m\\
S0421a & 68.3\textdegree & $  8.0 \pm   4.1$ & $   13-   22$ m\\
S0424c & 54.4\textdegree & $  7.8 \pm   4.1$ & $   13-   22$ m\\
S0475a & 23.9\textdegree & $  1.2 \pm   0.6$ & $    6-   12$ m\\
S0490a & 32.5\textdegree & $  1.3 \pm   0.7$ & $    6-   12$ m\\
S0500a & 22.5\textdegree & $  1.2 \pm   0.7$ & $    6-   11$ m\\
S0533a &  8.8\textdegree & $  2.3 \pm   1.2$ & $    7-   13$ m\\
S0534a & 72.2\textdegree & $ 14.2 \pm   6.1$ & $   17-   28$ m\\
S0542a & 40.2\textdegree & $  7.4 \pm   3.9$ & $   13-   22$ m\\
S0653a & 13.1\textdegree & $  1.1 \pm   0.5$ & $    6-   11$ m\\
S0661a & 13.8\textdegree & $  1.0 \pm   0.7$ & $    5-   11$ m\\
S0672a & 41.7\textdegree & $  4.3 \pm   2.3$ & $   10-   17$ m\\
S0712a & 14.4\textdegree & $  3.3 \pm   1.7$ & $    9-   15$ m\\
S0734a & 14.2\textdegree & $  6.5 \pm   3.4$ & $   11-   19$ m\\
S0756a & 35.6\textdegree & $ 35.0 \pm  15.1$ & $   23-   36$ m\\
S0758a & 28.7\textdegree & $  3.3 \pm   0.7$ & $   11-   16$ m\\
S0764a & 10.2\textdegree & $  2.6 \pm   1.7$ & $    7-   14$ m\\
S0774a & 23.0\textdegree & $  0.2 \pm   0.1$ & $    3-    7$ m\\
S0786a & 44.0\textdegree & $  0.9 \pm   0.7$ & $    6-   11$ m\\
S0792a & 12.9\textdegree & $  0.2 \pm   0.1$ & $    3-    6$ m\\
S0793a &  4.9\textdegree & $  0.5 \pm   0.4$ & $    3-    7$ m\\
S0794a & 29.3\textdegree & $ 16.2 \pm   8.4$ & $   17-   28$ m\\
S0799a & 41.0\textdegree & $  0.7 \pm   0.3$ & $    5-    9$ m\\
S0829a & 28.6\textdegree & $  4.7 \pm   2.5$ & $   10-   17$ m\\
S0842a &  9.0\textdegree & $  0.2 \pm   0.1$ & $    3-    6$ m\\
S0850a & 15.5\textdegree & $  2.2 \pm   1.1$ & $    7-   13$ m\\
S0860a & 11.5\textdegree & $  1.1 \pm   0.6$ & $    6-   11$ m\\
S0864b & 45.3\textdegree & $  5.5 \pm   2.9$ & $   11-   19$ m\\
S0869b & 56.8\textdegree & $ 11.4 \pm   5.9$ & $   15-   25$ m\\
S0879b & 11.0\textdegree & $  0.3 \pm   0.1$ & $    3-    7$ m\\
S0882a & 38.8\textdegree & $  5.3 \pm   2.3$ & $   11-   19$ m\\
S0889c & 21.9\textdegree & $  0.4 \pm   0.2$ & $    3-    7$ m\\
S0922a & 28.4\textdegree & $  0.4 \pm   0.2$ & $    3-    7$ m\\
S0923d & 90.4\textdegree & $ 12.0 \pm   4.0$ & $   15-   25$ m\\
S0923f & 33.2\textdegree & $  1.3 \pm   0.6$ & $    6-   12$ m\\
S0965a & 16.2\textdegree & $  1.0 \pm   0.5$ & $    6-   11$ m\\
S0976b & 29.3\textdegree & $129.0 \pm  90.1$ & $   34-   53$ m\\
S0981c &  5.7\textdegree & $  0.9 \pm   0.8$ & $    4-    9$ m\\
S0986c &  1.6\textdegree & $  0.3 \pm   0.2$ & $    2-    6$ m\\
S1000a & 126.9\textdegree & $1421.3 \pm 472.3$ & $  101-  148$ m\\
S1030a & 18.7\textdegree & $  0.3 \pm   0.2$ & $    3-    7$ m\\
S1032a & 17.4\textdegree & $  0.3 \pm   0.1$ & $    3-    7$ m\\
S1034a &  0.9\textdegree & $  1.9 \pm   1.4$ & $    6-   13$ m\\
S1047a & 25.2\textdegree & $  9.3 \pm   4.8$ & $   13-   22$ m\\
S1049d & 21.0\textdegree & $  3.9 \pm   2.0$ & $   10-   17$ m\\
S1090b & 18.1\textdegree & $  1.2 \pm   0.8$ & $    6-   11$ m\\
S1094b & 59.3\textdegree & $541.2 \pm 122.5$ & $   71-   97$ m\\
S1102a & 71.5\textdegree & $ 97.4 \pm  32.4$ & $   34-   53$ m\\
S1106a & 71.3\textdegree & $  1.6 \pm   0.4$ & $    7-   13$ m\\
S1143a & 16.0\textdegree & $  1.0 \pm   0.5$ & $    6-   11$ m\\
S1153a & 81.9\textdegree & $ 13.7 \pm   4.6$ & $   17-   28$ m\\
S1160a &  1.1\textdegree & $  0.4 \pm   0.2$ & $    3-    7$ m\\
S1171b & 19.8\textdegree & $  0.4 \pm   0.2$ & $    4-    8$ m\\
S1222a & 37.0\textdegree & $8306.9 \pm 4350.2$ & $  198-  281$ m\\
S1223a & 39.9\textdegree & $ 15.9 \pm   8.3$ & $   17-   28$ m\\
S1237a & 87.1\textdegree & $ 11.0 \pm   5.7$ & $   15-   25$ m\\
S1337a & 34.5\textdegree & $ 63.6 \pm  14.4$ & $   32-   45$ m\\
S1361a & 10.6\textdegree & $  0.5 \pm   0.3$ & $    4-    8$ m\\
S1415a & 85.0\textdegree & $ 74.6 \pm  32.1$ & $   34-   53$ m\\
\end{longtable}

\clearpage

\include{LFBB_events_grouped_with_new_tectonic_classes}

\end{document}

%% file: LFBB_events_grouped_with_new_tectonic_classes.tex
\begin{longtable}{lcccc}
\caption{Phlegra Montes LF/BB events.} \label{tab:phlegra_lfbb} \\
\toprule
Event & $\Delta$ (deg) & $M_0$ ($10^{12}$ N m) & $M_w$ & $f_c$ (Hz) \\
\midrule
\endfirsthead
\caption[]{Phlegra Montes LF/BB events.} \\
\toprule
Event & $\Delta$ (deg) & $M_0$ ($10^{12}$ N m) & $M_w$ & $f_c$ (Hz) \\
\midrule
\endhead
\midrule
\multicolumn{5}{r}{Continued on next page} \\
\midrule
\endfoot
\bottomrule
\endlastfoot
S1157a (B) & $35.6$ & $669.2 \pm 222.4$ & $3.82 \pm 0.10$ & 0.389 \\
S1157b (C) & $35.0$ & $1026.5 \pm 533.2$ & $3.94 \pm 0.15$ & 0.2 \\
S1157c (C) & $35.0$ & $204.8 \pm 106.4$ & $3.47 \pm 0.15$ & 0.4 \\
S1157e (C) & $34.5$ & $728.1 \pm 241.9$ & $3.84 \pm 0.10$ & 0.25 \\
S1157f (C) & $35.9$ & $2039.2 \pm 1059.3$ & $4.14 \pm 0.15$ & 0.2 \\
S1157g (B) & $35.9$ & $333.3 \pm 173.1$ & $3.62 \pm 0.15$ & 0.45 \\
\end{longtable}

\begin{longtable}{lcccc}
\caption{Valles Marineris LF/BB events.} \label{tab:valles_marineris_lfbb} \\
\toprule
Event & $\Delta$ (deg) & $M_0$ ($10^{12}$ N m) & $M_w$ & $f_c$ (Hz) \\
\midrule
\endfirsthead
\caption[]{Valles Marineris LF/BB events.} \\
\toprule
Event & $\Delta$ (deg) & $M_0$ ($10^{12}$ N m) & $M_w$ & $f_c$ (Hz) \\
\midrule
\endhead
\midrule
\multicolumn{5}{r}{Continued on next page} \\
\midrule
\endfoot
\bottomrule
\endlastfoot
S0976a (A) & $144.8$ & $2054.4 \pm 235.5$ & $4.14 \pm 0.03$ & 0.3 \\
\end{longtable}

\begin{longtable}{lcccc}
\caption{Amazonis LF/BB events.} \label{tab:amazonis_lfbb} \\
\toprule
Event & $\Delta$ (deg) & $M_0$ ($10^{12}$ N m) & $M_w$ & $f_c$ (Hz) \\
\midrule
\endfirsthead
\caption[]{Amazonis LF/BB events.} \\
\toprule
Event & $\Delta$ (deg) & $M_0$ ($10^{12}$ N m) & $M_w$ & $f_c$ (Hz) \\
\midrule
\endhead
\midrule
\multicolumn{5}{r}{Continued on next page} \\
\midrule
\endfoot
\bottomrule
\endlastfoot
S0183a (B) & $46.0$ & $34.4 \pm 11.4$ & $2.96 \pm 0.10$ & 0.3 \\
S0185a (B) & $60.4$ & $61.8 \pm 14.0$ & $3.13 \pm 0.07$ & 0.4 \\
S0325a (B) & $39.7$ & $506.1 \pm 114.5$ & $3.74 \pm 0.07$ & 0.25 \\
S0899d (B) & $50.0$ & $72.3 \pm 16.4$ & $3.17 \pm 0.07$ & 0.35 \\
S1097a (B) & $29.6$ & $76.3 \pm 17.3$ & $3.19 \pm 0.07$ & 0.57 \\
\end{longtable}

\begin{longtable}{lcccc}
\caption{Cerberus Fossae LF/BB events.} \label{tab:cerberus_lfbb} \\
\toprule
Event & $\Delta$ (deg) & $M_0$ ($10^{12}$ N m) & $M_w$ & $f_c$ (Hz) \\
\midrule
\endfirsthead
\caption[]{Cerberus Fossae LF/BB events.} \\
\toprule
Event & $\Delta$ (deg) & $M_0$ ($10^{12}$ N m) & $M_w$ & $f_c$ (Hz) \\
\midrule
\endhead
\midrule
\multicolumn{5}{r}{Continued on next page} \\
\midrule
\endfoot
\bottomrule
\endlastfoot
S0105a (B) & $32.5$ & $84.0 \pm 43.7$ & $3.22 \pm 0.15$ & 0.331 \\
S0133a (B) & $27.5$ & $16.4 \pm 5.4$ & $2.74 \pm 0.10$ & 0.35 \\
S0133b (B) & $27.3$ & $16.2 \pm 5.4$ & $2.74 \pm 0.10$ & 0.3 \\
S0173a (A) & $30.0$ & $477.6 \pm 54.7$ & $3.72 \pm 0.03$ & 0.3 \\
S0189a (B) & $33.6$ & $15.9 \pm 5.3$ & $2.73 \pm 0.10$ & 0.7 \\
S0190a (C) & $34.3$ & $16.0 \pm 8.3$ & $2.74 \pm 0.15$ & 0.3 \\
S0235b (A) & $28.9$ & $328.7 \pm 56.2$ & $3.61 \pm 0.05$ & 0.4 \\
S0235c (C) & $29.1$ & $13.5 \pm 7.0$ & $2.69 \pm 0.15$ & 0.35 \\
S0235e (C) & $30.4$ & $14.1 \pm 7.3$ & $2.70 \pm 0.15$ & 0.3 \\
S0290b (B) & $30.4$ & $48.6 \pm 16.2$ & $3.06 \pm 0.10$ & 0.44 \\
S0323d (C) & $29.1$ & $67.8 \pm 35.2$ & $3.15 \pm 0.15$ & 0.12 \\
S0327d (C) & $26.8$ & $35.1 \pm 18.2$ & $2.96 \pm 0.15$ & 0.35 \\
S0345e (C) & $28.6$ & $23.6 \pm 12.3$ & $2.85 \pm 0.15$ & 0.3 \\
S0407a (B) & $29.3$ & $26.7 \pm 6.0$ & $2.88 \pm 0.07$ & 0.4 \\
S0409d (B) & $31.3$ & $58.8 \pm 19.5$ & $3.11 \pm 0.10$ & 0.3 \\
S0484b (B) & $29.8$ & $27.2 \pm 6.2$ & $2.89 \pm 0.07$ & 0.631 \\
S0748a (C) & $29.1$ & $24.1 \pm 12.5$ & $2.85 \pm 0.15$ & 0.3 \\
S0784a (B) & $34.5$ & $81.7 \pm 27.1$ & $3.21 \pm 0.10$ & 0.5 \\
S0802a (B) & $30.0$ & $39.3 \pm 11.0$ & $3.00 \pm 0.08$ & 0.7 \\
S0802b (C) & $29.3$ & $27.2 \pm 14.1$ & $2.89 \pm 0.15$ & 0.35 \\
S0809a (A) & $29.8$ & $153.1 \pm 34.6$ & $3.39 \pm 0.07$ & 0.3 \\
S0809b (A) & $29.8$ & $140.9 \pm 46.8$ & $3.37 \pm 0.10$ & 0.25 \\
S0820a (A) & $30.2$ & $109.7 \pm 24.8$ & $3.29 \pm 0.07$ & 0.35 \\
S0820b (B) & $31.4$ & $11.4 \pm 4.3$ & $2.64 \pm 0.11$ & 0.4 \\
S0833a (C) & $34.7$ & $20.3 \pm 10.5$ & $2.80 \pm 0.15$ & 0.3 \\
S0850c (C) & $32.2$ & $13.5 \pm 4.5$ & $2.69 \pm 0.10$ & 0.3 \\
S0861a (C) & $32.2$ & $42.1 \pm 21.9$ & $3.02 \pm 0.15$ & 0.2 \\
S0864a (A) & $29.6$ & $67.2 \pm 15.2$ & $3.15 \pm 0.07$ & 0.776 \\
S0864e (C) & $31.6$ & $92.7 \pm 48.2$ & $3.24 \pm 0.15$ & 0.3 \\
S0916d (B) & $29.3$ & $31.0 \pm 10.3$ & $2.93 \pm 0.10$ & 0.7 \\
S0918a (B) & $27.3$ & $64.8 \pm 14.7$ & $3.14 \pm 0.07$ & 0.35 \\
S0918b (C) & $25.5$ & $26.5 \pm 13.8$ & $2.88 \pm 0.15$ & 0.25 \\
S0923a (C) & $33.2$ & $10.9 \pm 5.7$ & $2.63 \pm 0.15$ & 0.6 \\
S0943h (C) & $26.1$ & $48.2 \pm 25.0$ & $3.06 \pm 0.15$ & 0.3 \\
S0982b (C) & $30.0$ & $81.2 \pm 42.2$ & $3.21 \pm 0.15$ & 0.17 \\
S1000b (B) & $25.2$ & $58.6 \pm 30.4$ & $3.11 \pm 0.15$ & 0.3 \\
S1015f (A) & $27.5$ & $26.1 \pm 3.0$ & $2.88 \pm 0.03$ & 0.4 \\
S1015g (B) & $28.4$ & $18.3 \pm 7.0$ & $2.77 \pm 0.11$ & 0.4 \\
S1016e (B) & $29.3$ & $26.6 \pm 10.2$ & $2.88 \pm 0.11$ & 0.2 \\
S1022a (A) & $30.7$ & $409.1 \pm 135.9$ & $3.67 \pm 0.10$ & 0.175 \\
S1039b (B) & $30.5$ & $31.3 \pm 7.1$ & $2.93 \pm 0.07$ & 0.5 \\
S1048d (A) & $30.2$ & $358.2 \pm 119.0$ & $3.64 \pm 0.10$ & 0.1 \\
S1197a (B) & $32.0$ & $450.2 \pm 193.8$ & $3.70 \pm 0.12$ & 0.8 \\
\end{longtable}

\begin{longtable}{lcccc}
\caption{Other LF/BB events.} \label{tab:other_lfbb} \\
\toprule
Event & $\Delta$ (deg) & $M_0$ ($10^{12}$ N m) & $M_w$ & $f_c$ (Hz) \\
\midrule
\endfirsthead
\caption[]{Other LF/BB events.} \\
\toprule
Event & $\Delta$ (deg) & $M_0$ ($10^{12}$ N m) & $M_w$ & $f_c$ (Hz) \\
\midrule
\endhead
\midrule
\multicolumn{5}{r}{Continued on next page} \\
\midrule
\endfoot
\bottomrule
\endlastfoot
S0152a (C) & $50.0$ & $26.1 \pm 13.5$ & $2.88 \pm 0.15$ & 0.4 \\
S0154a (C) & $50.0$ & $73.5 \pm 38.2$ & $3.18 \pm 0.15$ & 0.25 \\
S0167a (C) & $50.0$ & $164.5 \pm 85.5$ & $3.41 \pm 0.15$ & 0.15 \\
S0205a (C) & $50.0$ & $29.3 \pm 15.2$ & $2.91 \pm 0.15$ & 0.25 \\
S0217a (C) & $50.0$ & $26.1 \pm 13.5$ & $2.88 \pm 0.15$ & 0.3 \\
S0226b (C) & $50.0$ & $73.5 \pm 38.2$ & $3.18 \pm 0.15$ & 0.17 \\
S0234c (C) & $50.0$ & $29.3 \pm 15.2$ & $2.91 \pm 0.15$ & 0.3 \\
S0320b (C) & $35.8$ & $84.7 \pm 28.1$ & $3.22 \pm 0.10$ & 0.16 \\
S0345a (C) & $50.0$ & $65.5 \pm 34.0$ & $3.14 \pm 0.15$ & 0.2 \\
S0357a (C) & $50.0$ & $58.4 \pm 30.3$ & $3.11 \pm 0.15$ & 0.15 \\
S0362b (C) & $50.0$ & $46.4 \pm 24.1$ & $3.04 \pm 0.15$ & 0.3 \\
S0377c (C) & $50.0$ & $73.5 \pm 38.2$ & $3.18 \pm 0.15$ & 0.18 \\
S0405c (C) & $50.0$ & $18.5 \pm 9.6$ & $2.78 \pm 0.15$ & 0.35 \\
S0421b (C) & $50.0$ & $52.0 \pm 27.0$ & $3.08 \pm 0.15$ & 0.2 \\
S0423a (C) & $50.0$ & $36.8 \pm 19.1$ & $2.98 \pm 0.15$ & 0.2 \\
S0474a (B) & $37.7$ & $9.7 \pm 3.7$ & $2.59 \pm 0.11$ & 0.5 \\
S0784b (C) & $42.8$ & $50.8 \pm 16.9$ & $3.07 \pm 0.10$ & 0.3 \\
S0832a (C) & $50.0$ & $41.3 \pm 21.5$ & $3.01 \pm 0.15$ & 0.15 \\
S0892b (C) & $50.0$ & $92.5 \pm 48.1$ & $3.24 \pm 0.15$ & 0.2 \\
S0936b (C) & $50.0$ & $16.5 \pm 8.5$ & $2.74 \pm 0.15$ & 0.3 \\
S1012d (B) & $38.1$ & $75.9 \pm 25.2$ & $3.19 \pm 0.10$ & 0.3 \\
S1032b (C) & $50.0$ & $34.7 \pm 13.0$ & $2.96 \pm 0.11$ & 0.3 \\
S1038a (C) & $50.0$ & $232.4 \pm 120.7$ & $3.51 \pm 0.15$ & 0.3 \\
\end{longtable}